\documentclass[reprint, superscriptaddress, nofootinbib, eqsecnum, longbibliography, amsmath, amssymb, aps, showkeys]{revtex4-1}

\usepackage[dvipsnames]{xcolor}
\usepackage{graphicx}
\usepackage{dcolumn}
\usepackage{bm}
\usepackage[colorlinks]{hyperref}
\usepackage[capitalise]{cleveref}
\usepackage{tabularx}
\usepackage{xspace}
\usepackage{xcolor}
\usepackage{amssymb,amsmath,bm}

\usepackage[normalem]{ulem}
\usepackage[utf8]{inputenc}

\usepackage[inline]{enumitem}

\newcommand\myshade{80}
\colorlet{mylinkcolor}{ForestGreen}
\colorlet{mycitecolor}{Red}
\colorlet{myurlcolor}{violet}
\usepackage[dvipsnames]{xcolor}

\hypersetup{
  linkcolor  = mylinkcolor!\myshade!black,
  citecolor  = mycitecolor!\myshade!black,
  urlcolor   = myurlcolor!\myshade!black,
  colorlinks = true
}

\newcommand{\calE}{\mathcal{E}}
\newcommand{\rmsp}{\mathrm{sp}}
\newcommand{\rmDM}{\mathrm{DM}}
\newcommand{\rmIn}{\mathrm{in}}
\newcommand{\avg}[1]{\left\langle #1 \right\rangle}

\DeclareSymbolFont{mathtx}{OML}{txmi}{m}{it}
\DeclareMathAlphabet\mathbfcal{OMS}{cmsy}{b}{n}
\DeclareMathSymbol{v}{\mathalpha}{mathtx}{118}

\begin{document}

\newcommand{\MyEmptyAffiliation}{ $:')$ }
\newcommand{\GRAPPA}{Gravitation Astroparticle Physics Amsterdam (GRAPPA),\\ Institute for Theoretical Physics Amsterdam and Delta Institute for Theoretical Physics,\\ University of Amsterdam, Science Park 904, 1098 XH Amsterdam, The Netherlands}
\newcommand{\UdeM}{Département de Physique, Université de Montréal, 1375 Avenue Thérèse-Lavoie-Roux, Montréal, QC H2V 0B3, Canada}
\newcommand{\Mila}{Mila -- Quebec AI Institute, 6666 St-Urbain, \#200, Montreal, QC, H2S 3H1}
\newcommand{\IFCA}{Instituto de F\'isica de Cantabria (IFCA, UC-CSIC), Av.~de Los Castros s/n, 39005 Santander, Spain}

\preprint{}

\title{Sharpening the dark matter signature in gravitational waveforms I:\\Accretion and eccentricity evolution}

\author{Theophanes K. Karydas}
\email{tk.karydas@gmail.com}
\affiliation{\GRAPPA}
\author{Bradley J. Kavanagh}
\email{kavanagh@ifca.unican.es}
\affiliation{\IFCA}
\author{Gianfranco Bertone}
\email{g.bertone@uva.nl}
\affiliation{\GRAPPA}

\begin{abstract}
Dark matter overdensities around black holes can alter the dynamical evolution of a companion object orbiting around it, and cause a dephasing of the gravitational waveform. Here, we present a refined calculation of the co-evolution of the binary and the dark matter distribution, taking into account the accretion of dark matter particles on the companion black hole, and generalizing previous quasi-circular calculations to the general case of eccentric orbits. These calculations are validated by dedicated $N$-body simulations. We show that accretion can lead to a large dephasing, and therefore cannot be neglected in general. We also demonstrate that dark matter spikes tend to circularize eccentric orbits faster than previously thought.
\end{abstract}


\maketitle

\setcounter{tocdepth}{2}
\setcounter{secnumdepth}{2}

\section{Introduction}

The next generation of space-based gravitational wave (GW) detectors, such as LISA \cite{LISA} and Taiji \cite{Hu:2017mde}, will open a novel era for gravitational wave astronomy through their substantially lower frequency range compared to current detectors like LIGO \cite{LIGO}, Virgo \cite{VIRGO}, and KAGRA \cite{kagracollaboration2020overview}. These instruments will allow the observation of long duration waveforms, for up to several years, enabling precision studies of intermediate- and extreme- mass ratio inspirals (IMRIs and EMRIs, respectively). 
EMRI and IMRI waveforms are a powerful probe of black hole environments \cite{Macedo:2013qea,Barausse:2014tra,Barausse:2014pra,Zwick:2022dih,Cole:2022yzw,daghigh2023effect,Cardoso:2019rou,Baumann_2022,Bamber_2023,Aurrekoetxea:2023jwk}, making it possible in particular to detect the presence around black holes of large dark matter (DM) overdensities, commonly referred to as dark matter `spikes' \cite{Gondolo:1999ef,Ullio:2001fb, Zhao:2005zr,Bertone:2005xz,Hannuksela_2020}. Dark matter is expected in fact to imprint a characteristic dephasing on the waveform due to dynamical friction \cite{Eda_2013}.

\citet{Kavanagh:2020cfn} demonstrated that the distribution of dark matter is strongly affected by the interaction with the companion, and they developed a  model to co-evolve the binary and the dark matter distribution. Once the feedback on the dark matter distribution is included, the dephasing induced by dynamical friction is largely suppressed compared to the static case explored widely in the literature~\cite{Eda_2015,Yue:2017iwc,Yue:2018vtk,Edwards:2019tzf,Zhang:2024ugv,Speeney:2022ryg,Montalvo:2024iwq}. Even so, the DM dephasing effect in quasi-circular inspirals may still be large enough to be detectable with future detectors~\cite{Coogan:2021uqv,Cole:2022ucw}.
\citet{Yue_2019} investigated how dynamical friction impacts a binary's eccentricity using a toy model based on dimensional analysis, and showed that it leads to a fast eccentrification of the orbit. However, their spike lacked a phase-space distribution, and its particles had no relative velocity with respect to the orbiting companion. Despite this being an adequate approximation for extreme-mass-ratio, quasi-circular inspirals in absence of feedback, the velocity dependence of dynamical friction complicates the calculation of the eccentricity evolution. This effect has recently been explored by \citet{Becker_2022} and \citet{Li2}, where it was found that an isotropic spike induces an additional circularization instead.

Here, we build on the results of Ref.~\cite{Kavanagh:2020cfn}, and we present two new contributions to the characterization of compact object binaries evolving in presence of a dark matter spikes: the accretion of particles onto the companion object, and a detailed description of the evolution and impact of orbital eccentricity. We derive a semi-analytical model for particle accretion onto the orbiting companion object to describe the binary's backreaction and the spike's depletion. This generalizes the formalism for accretion recently proposed in Ref.~\cite{PhysRevD.108.124062}. In addition, we extend the analysis beyond quasi-circular inspirals by incorporating orbital eccentricity to the feedback formalism developed in Ref.~\cite{Kavanagh:2020cfn}. Finally, we explore the impact of both accretion and eccentricity -- including the effects of feedback -- on the GW dephasing induced by the DM spike.

In a companion paper~\cite{BetterSpikesII} (hereafter Paper II), we present the $N$-body code \texttt{NbodyIMRI}~\cite{NbodyIMRI}, tailored to model the evolution of IMRIs in DM spikes. In Paper II, we use the code to study the strength of the dynamical friction force (and the corresponding feedback) in these systems. Here, we use the code to verify the strength of the force induced by accretion and to validate our formalism for accretion-induced feedback.

The layout of this work is as follows.
\cref{sec:intro} introduces dark matter spikes and outlines the formalism for binary system evolution in non-vacuum space, extended to accommodate orbital eccentricity and the varying mass of the secondary.
\cref{sec:accretion} discusses isotropic particle accretion onto the companion object and the subsequent spike depletion.
In \cref{sec:n_bodies}, we present the results of $N$-body simulations, described in Paper II, which enable us to validate our model for particle accretion by the secondary black hole.
Finally, \cref{sec:circ_acc,sec:eccentricity} presents numerical results on spike and binary system evolution, followed by concluding remarks in \cref{sec:conclusion}.

\section{Inspirals in dark matter spikes}
\label{sec:intro}
\begin{figure}[tb]
    \includegraphics[width=0.85\columnwidth]{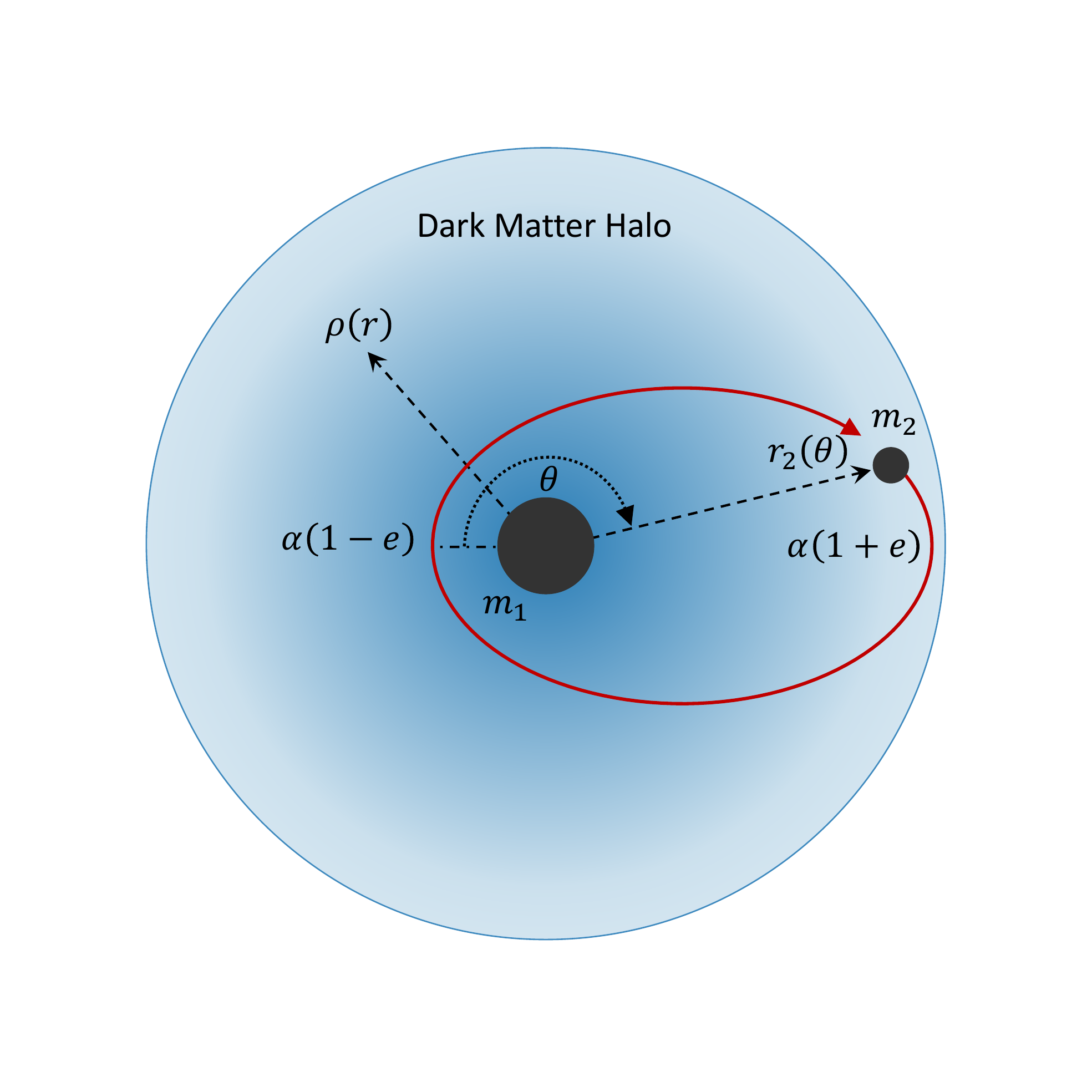}
    \caption{\textbf{An intermediate mass ratio inspiral embedded within a dark matter spike.}
    A black hole of mass $m_2$ orbits around a more massive black hole with mass $m_1 \gg m_2$, surrounded by dark matter. The orbit is characterised by the semi-major axis $a$ and orbital eccentricity $e$.
    \label{fig:binary_spike_circ}
    }
\end{figure}

Building on our previous work \cite{Kavanagh:2020cfn}, we consider a small mass ratio binary, consisting of a central, stationary black hole $m_1$, and an orbiting black hole with mass $m_2$, with mass ratio $q \equiv m_2/m_1 \leq 10^{-2.5}$ \cite{Kavanagh:2020cfn}. We further assume that the central black hole is surrounded by an isotropic, spherically symmetric distribution of non-interacting dark matter particles that is centered on $m_1$ as depicted in \cref{fig:binary_spike_circ}.

\subsection{The cold dark matter spike} \label{sec:spike}

We assume a `spike' of cold dark matter formed by the adiabatic growth of the central black hole \cite{Gondolo:1999ef}. 
The density distribution of the spike can be modeled as the following power law profile\footnote{The dark matter density profile around a black hole is  generally sensitive to the particle properties of dark matter \cite{Hannuksela_2020,Alvarez_2021,Shapiro_2016,PhysRevD.107.083005} and the surrounding stellar population~\cite{Bertone:2005hw,Shapiro_2022}, which may lead to the formation of cores or other scale breaks in the power law.}:
\begin{equation} \label{eq:rhoDM}
    \rho_\rmDM(r) = \left\{
    \begin{array}{ll}
      \rho_6 \left(\frac{r}{r_6}\right)^{-\gamma_\rmsp}   &  r_\rmIn \leq r \leq r_\rmsp\\
       0.  & r < r_\rmIn
    \end{array}
    \right. \, ,
\end{equation}
where $r$ is the distance from the center of the spike, $\rho_6$ is the density normalization at the fixed distance \linebreak $r_6 = 10^{-6} \,\mathrm{pc}$, and $r_\rmsp$ is its length.
Accretion onto the central black hole severely depletes the inner spike at radius $r_\mathrm{in} = 2R_s = 4 G m_1/c^2$ \cite{Sadeghian_2013,Ferrer_2017}. This radius is much smaller than the scales we are interested in, and in what follows we adopt a power-law to describe the density profile.
The slope of the spike will be in the range $\gamma_\rmsp \in [2.25, 2.5]$ when forming around astrophysical black holes and $\gamma_\rmsp = 2.25$ around primordial black holes \cite{Mack:2006gz,Eroshenko:2016yve,Adamek_2019,Eroshenko_2020,Boudaud:2021irr}.

Alternatively, the halo can be described through its phase space distribution function $f = m_\mathrm{DM} \mathrm{d} N/\mathrm{d}^3 \mathbf{r}\,\mathrm{d}^3\mathbf{v}$, which additionally encodes information about the orbital properties of the particles. For spherically symmetric and isotropic profiles this function is only described by the relative specific energy of the particles, hence we can write
\begin{equation}
    f \to f(\calE) \qquad \text{with} \qquad
    \calE(r,v) = \Psi(r) - \frac{1}{2}v^2\,,
\end{equation}
where the energy is defined such that bound particles have $\calE > 0$. Here, $\Psi(r) = \Phi_0 - \Phi(r)$ is the total relative potential from all sources of gravity, where we take the reference potential $\Phi_0$ to be zero. For the systems we are interested in, the gravitational potential due to the DM spike will typically be sub-dominant to that of the central massive BH. Therefore, inside the spike we consider only the potential of the central BH, giving $\Psi(r) = G m_1/r$, and thus we ignore the precession to the orbit that may be generated by the spike's potential~\cite{Dai:2021olt}.\footnote{We have verified explicitly that for the spike parameters we consider in this work, the effect on the dynamics due to the spike's potential is 3-5 orders of magnitude smaller than other spike-induced effects from accretion and dynamical friction.}

The distribution function of an isotropic density profile is constructed through the Eddington inversion procedure (see p. 290, \cite{BinneyAndTremaine}), and the mass density is reclaimed by averaging over all possible particle velocities. We follow the time-evolution of the density by studying $f(\calE, t)$, similarly to the work in Refs.~\cite{Bertone:2005hw,PhysRevD.75.043517,Kavanagh:2020cfn}.

\subsection{Orbital and Environmental evolution equations}

We describe the state of the binary using its semi-major axis ($a$) and eccentricity ($e$) which will be evolved in time through constructed diferential equations that account for environmental effects and gravitational wave radiation. To do that, we operate in the post-Newtonian approximation where ignoring the first two conservative PN terms, we write the Newtonian energy,
\begin{align}
    \label{eq:energy_1}
      E &= \frac{1}{2} m_1 u_1^2 +\frac{1}{2} m_2 u_2^2 + V(|\bm{r_2}-\bm{r_1}|) \\ &= \frac{1}{2} \mu u^2 - \frac{G \mu m}{r} = -\frac{Gm\mu}{2a}\,, \label{eq:energy_2}
  \end{align}
and angular momentum,
\begin{align}
    \label{eq:angtum}
      L &= ||m_1 \bm{r}_1 \times \bm{u}_1 +m_2 \bm{r}_2 \times \bm{u}_2|| \\
      & = \mu \sqrt{G m a (1-e^2)}\,,
  \end{align}
where $\bm{r}_i$ are the positions of the two BHs with respect to the center of mass, $\bm{u}_i$ the orbital velocities of the two components, $m_i$ their masses, and $V(r)$ the gravitational potential. The quantities $\mu = m_1 m_2 /m$ and $m = m_1+~\!m_2$ are the reduced and total mass, respectively. The potential is given by $V(r) = - G m_1 m_2/r$, where $r = |\bm{r}_2 -\bm{r}_1|$ is the separation of the binary, and $u$ is the orbital velocity of the reduced particle.

We illustrate one such orbit in \cref{fig:binary_spike_circ}, where the separation and orbital velocity can be written as functions of the true anomaly $\theta$:
\begin{align} \label{eq:separation}
  r(\theta) &= \frac{a (1 -e^2)}{1 +e\cos\theta} \,,\\
  u^2(\theta) &= G m \bigg(\frac{2}{r} -\frac{1}{a} \bigg) \,.
\end{align}
This contribution vanishes for circular orbits at $e=0$ where $r \equiv a$ and $u^2=Gm/r$. The orbital period simply depends on the semi-major axis alone
\begin{equation}
    T = 2\pi \sqrt{\frac{a^3}{Gm}}\,. \label{eq:period}
\end{equation}
We will assume throughout that the mass ratio $q$ is sufficiently small that we can write $r_2 \approx r$ (i.e.~that $m_1$ lies at the centre of mass). 

During the inspiral, energy and angular momentum are carried away from the binary by gravitational waves and dark matter particles, altering its orbit.
\begin{align}
    \label{eq:averageLoss}
    \dot{E} &= \dot{E}_\mathrm{GW} + \dot{E}_\mathrm{env}\,, \\
    \dot{L} &= \dot{L}_\mathrm{GW} + \dot{L}_\mathrm{env}\,.
  \end{align}
By differentiation of the energy and angular momentum \cref{eq:energy_1,eq:energy_2}, we can derive the evolution of the semi-major axis and eccentricity, respectively
\begin{equation}
  \label{eq:da_dtdedt}
    \dot{a} = -a \frac{\dot{E}}{E} \,,\qquad \dot{e} = -\frac{1-e^2}{e} \bigg( \frac{\dot{E}}{2E} +\frac{\dot{L}}{L} \bigg)\,.
\end{equation}

\paragraph*{\bf Gravitational radiation.}
Gravitational wave emission is the dominant source of energy and angular momentum loss for a dressed binary. For non-spinning, point-like black holes, the energy and angular momentum loss rates of a binary are taken at their first order, the 2.5 PN order, which are~\cite{Maggiore:2007ulw}:
\begin{align}
  \label{eq:GWlosses}
    \dot{E}_\mathrm{GW} &= -\frac{32 G^4}{5c^5} \frac{\mu^2 m^3}{a^5}\frac{1 +\frac{73}{24}e^2 +\frac{37}{96}e^4 }{(1-e^2)^{7/2}}\,, \\
    \dot{L}_\mathrm{GW} &= -\frac{32 G^{7/2}}{5c^5} \frac{\mu^2 m^{5/2}}{a^{7/2}}\frac{1 +\frac{7}{8}e^2 }{(1-e^2)^{2}}\,.
\end{align}

\paragraph*{\bf Environment-induced forces.}
Effects from the binary's environment can take the form of a force or an exchange of mass (which results not only in an increase in the mass of the secondary, but also in an effective force due to the accretion of momentum). For isotropic DM distributions, the force $F$ generally acts on the direction of the companion's motion and generates the following instantaneous energy and angular momentum loss rates:
\begin{equation}
  \label{eq:forceLosses}
  \frac{\mathrm{d}E}{\mathrm{d}t} = u F \,, \qquad \frac{\mathrm{d}L}{\mathrm{d}t} = \sqrt{G m a(1 -e^2)} \ \frac{F}{u}\,.
\end{equation}

\paragraph*{\bf Dynamical Friction.}
\label{sec:dynamicalfriction}
Dynamical friction (DF) \cite{1943ApJ....97..255C,chandra2,chandra3} is the accumulated deceleration on a test body due to gravitational scattering with a particle distribution of mass density $\rho_\mathrm{DM}$. In the context of this work, the orbiting companion slingshots DM particles around its path and steadily loses energy by displacing them towards outer orbits. When the companion of mass $m_2$ is moving with orbital velocity $u$, the force is written as
\begin{equation} \label{eq:DF}
    \bm{F}_\mathrm{DF} = -4\pi G^2 m_2^2 \frac{{\rho_\mathrm{DM}}}{u^2} \mathbfcal{C}_\mathrm{DF} \,.
\end{equation}
Here, $\mathbfcal{C}_\mathrm{DF}$ is generally a dimensionless vector that controls the direction and magnitude of the force, being sensitive to the particle distribution and geometry of the system. In principle, when $\mathbfcal{C}_\mathrm{DF} \cdot \mathbf{u} > 0$ the accumulated effect no longer dissipates energy and relates to ``dynamical heating'' \cite{BinneyAndTremaine}.

For an isotropic background, Chandrasekhar's formula describes a dissipative force that acts in the direction of the secondary's motion, and has $\mathcal{C}_\mathrm{DF}(u) = \ln\Lambda \, \xi_{<u}$, where $\xi_{<u}$ is the fraction of particles moving slower than the orbital velocity of the test body and $\ln\Lambda$ acts as ``Coulomb logarithm'' to the force, defined as~\cite{BinneyAndTremaine}:
\begin{equation} \label{eq:Lambda}
  \ln \Lambda = \ln \sqrt{\frac{b^2_\mathrm{max} +b^2_{90}}{b^2_\mathrm{min} +b^2_{90}}}\,.
\end{equation}
Here, $b_\mathrm{max}$ and $b_\mathrm{min}$ are the maximum and minimum impact parameters for which the two-body encounters contributing to the force are effective,\footnote{
  We set $b_\mathrm{max} = \sqrt{\frac{m_2}{m_1}} r_2$ following Ref.~\cite{Kavanagh:2020cfn} and $b_\mathrm{min} = b_\mathrm{acc}$ based on our discussion in \cref{sec:accretion}. We note that numerical studies in Paper II~\cite{BetterSpikesII} suggest that a more suitable parametrization may be $b_\mathrm{max}\approx 0.3\,r_2$. 
  We note that both of these definitions differ from previous works in the literature where eccentricity is treated. \citet{Yue_2019} implemented the constant logarithm $\ln\Lambda = 10$ based on the work of \citet{Amaro_Seoane_2007} on extreme mass ratio inspirals inside stellar distributions, and \citet{Becker_2022} used $\Lambda \approx \sqrt{m_1/m_2}$, which is derived as ours, but, in the quasi-circular regime.
  } while $b_{90} = Gm_2 /u^2$ is the impact parameter which gives a $90^\circ$ deflection. An extension of the standard Chandrasekhar formula was recently described in Ref.~\cite{Dosopoulou:2023umg}. Indeed, the $N$-body results presented in Paper II~\cite{BetterSpikesII} suggest that Chandrasekhar's formula may slightly overestimate the dynamical friction force. However, for simplicity, we will parametrize the strength of the force as $\mathcal{C}_\mathrm{DF}(u) = \ln\Lambda \, \xi_{<u}$, as described above.\\

\paragraph*{\bf Dynamic spikes -- HaloFeedback.}

As the companion moves around the central black hole, dark matter particles scatter around it, and are redistributed to new orbits via dynamical friction \textit{feedback}~\cite{Kavanagh:2020cfn}. The change in the spike can be described through the change in its distribution function $\dot{f}_{DF}(\calE)$, described by the following integro-differential equation:
\begin{equation} \label{eq:df_dt}
  \dot{f}_{DF}(\calE) =  \frac{\Delta_{+}f(\calE) - p_\mathrm{DF}(\calE)f(\calE) }{T} \,,
\end{equation}
The total probability that a particle with energy $\calE$ will scatter with the companion during a single orbit is written as:
\begin{equation}
    p_\mathrm{DF}(\calE) = \int P_{\calE}(\Delta\calE) \,\mathrm{d}\Delta\calE\,.
\end{equation}
and $\Delta_{+}f(\calE)$ is the change in the distribution function due to the scattering of particles from $\calE - \Delta \calE \rightarrow \calE$:
\begin{align}
  \Delta_{+}f(\calE) &= \int \bigg( \frac{\calE}{\calE -\Delta\calE} \bigg)^{5/2} \!\!\!\!\!\! f(\calE -\Delta\calE) P_{\calE -\Delta\calE}(\Delta \calE) \,\mathrm{d}\Delta\calE\,,
\end{align}
Here, $P_{\calE}(\Delta\calE)$ the probability of a particle with energy $\calE$ to scatter with the companion and gain energy $\Delta\calE$. The integration limits are over the minimum and maximum energy transfer $\Delta\calE_\mathrm{min,\,max}$ that can be imparted to a particle with energy $\calE$, calculated from $\Delta \calE(b)~=~-2u^2 (1+b^2/b_{90}^2)^{-1}$ for the maximum and minimum impact parameters $b_\mathrm{max,min}$ respectively.
For a black hole in a circular inspiral $P_{\calE}(\Delta\calE)$ is known semi-analytically as
\begin{equation}
  P_\calE(\Delta\calE) = \frac{16 m_2^2}{ \pi G m_1^3}\frac{r_2}{u^2}\frac{\calE^{5/2}}{\Delta\calE^2} \int{ \sqrt{\Psi' -\calE} \, \mathrm{d}\phi}\,,
\end{equation}
where $\Psi' = \Psi(r_2)(1 - b/r_2 \cos\phi)$. The final integration can be evaluated analytically at $O(b_\mathrm{max}/r_2)$ using special functions and an implementation of this feedback formalism is found in the publicly available code \href{https://github.com/bradkav/HaloFeedback}{https://github.com/bradkav/HaloFeedback}~\cite{HaloFeedback}.\\

\paragraph*{\bf Putting the framework together.}

In summary, in absence of mass accretion, we can describe the evolution of a black hole binary inside of a dark matter spike using \cref{eq:da_dtdedt} and by accounting for changes in its energy and angular momentum. We take these to be the gravitational wave radiation at the 2.5 PN order using \cref{eq:GWlosses}, and the total exchange of energy and angular momentum between the black hole companion and its surrounding particles. For example, by means of the dissipative force of dynamical friction as calculated with \cref{eq:forceLosses,eq:DF}. The backreaction (feedback) to the spike from this exchange is then accounted for using \cref{eq:df_dt}.

In the following section, we will further extend this framework by accounting for the mass accretion of particles onto the black hole companion. This is approached at Newtonian order while employing the absorption cross-section for small Schwarzschild black holes in general relativity \cite{PhysRevD.14.3251}, specifically in the low-velocity limit. Consequently, we demonstrate how the evolution of the orbital elements, as outlined in \cref{eq:da_dtdedt}, is modified to account for the increasing mass of the companion. Additionally, we introduce two new components: a dissipative force from the surrounding environment, denoted as $F_{acc}$, and a feedback mechanism that ensures mass conservation. Thus, binaries in our framework will be modeled entirely by these four differential equations,
\begin{equation}
  \begin{aligned}
    \dot{E}_\mathrm{tot} &= \dot{E}_{GW} +\dot{E}_{DF} +\dot{E}_{acc}\,,\\
    \dot{L}_\mathrm{tot} &= \dot{L}_{GW} +\dot{L}_{DF} +\dot{L}_{acc}\,,\\
    \dot{f}(\calE) &= \dot{f}_{DF}(\calE) +\dot{f}_{acc}(\calE)\,,\\
    \dot{m}_2 &= \dot{m}_\mathrm{acc}\,.
  \end{aligned}
\end{equation}
Before applying this system to the study of quasi-circular and elliptical inspirals in \cref{sec:circ_acc} and \cref{sec:eccentricity} respectively, we first employ the N-body algorithm from Paper II~\cite{BetterSpikesII} in \cref{sec:n_bodies} to validate the analytic formalism developed for accretion.

\section{Particle accretion onto the companion} \label{sec:accretion}

Accretion is the process that describes the inflow of particles towards a gravitating object. The exchange of energy and momentum between a distribution of collisionless particles and an attracting object gives rise to dynamical friction~\cite{1943ApJ....97..255C,chandra2,chandra3,BinneyAndTremaine}. If the object is a black hole, particles on sufficiently close trajectories will be absorbed (along with their momentum) by the BH~\cite{movingblack}. The accretion of gas onto black holes has also been explored in the literature \cite{Barausse_2008}.

Here, we present a formalism for accretion onto a body from any isotropic distribution of collisionless particles. In this section, will show how particles captured by an object influence its orbit and subsequently alter a binary inspiral. We will consider two effects: 
\begin{itemize}
\small
    \item \textbf{Orbital perturbation:} We study how the accretion of particles alters the inspiral of the binary. Not only does the transfer of mass increase the inertia of the companion~\cite{1992CeMDA..53..227P}, but the accumulating absorption of momentum amounts to a new force akin to dynamical friction.
    
    \item \textbf{Halo depletion:} We present a formalism for calculating the evolution of the DM spike in response to particles being removed from it.
\end{itemize}

\subsection{Accretion of mass and momentum}

Consider a large static distribution of constant density $\rho\,$, made up of particles with mass $\mu$, and an accreting body with mass $m$ traversing the distribution with an absorption cross-section $\sigma$. If the radius of the accreting object is larger than the mean free distance of the particles, we have to deal with cohesion forces~\cite{Giddings_2008}; however, we focus here on the collisionless case relevant for DM. A body which accretes particles at a radius $r_\mathrm{acc}$ has an absorption cross section:
\begin{equation} \label{eq:c_section}
  \sigma_N = \zeta \pi r_\mathrm{acc}^2 \bigg( 1 +\frac{r_s}{r_\mathrm{acc}}\frac{1}{\beta_0^2} \bigg)\,,
\end{equation}\\
where $r_s = 2Gm/c^2$ is the Schwarzschild radius associated with that mass.\footnote{We keep the accretion radius $r_\mathrm{acc}$ general, and potentially distinct from the Schwarzschild radius, in anticipation of the $N$-body simulations presented in \cref{sec:n_bodies}, where we employ an artificially large value of $r_\mathrm{acc}$.} The second term in brackets accounts for gravitational focusing with $\beta_0 = V_0/c$ and $V_0$ the relative velocity between the body and the particle being absorbed. The factor $\zeta$ depends on the nature of the body: $\zeta = 1$ for a Newtonian body and $\zeta = 4$ for a BH. This parameterization is chosen because at $\mathcal{O}(\beta_0^4)$, a small Schwarzschild black hole~\cite{PhysRevD.14.3251} will capture particles with a similar cross-section,
\begin{equation}
  \sigma_\mathrm{BH} = \frac{16 \pi G^2 m^2}{c^4}\bigg( 1 + \frac{1}{\beta_0^2} \bigg) \,.
\end{equation}

By dimensional analysis, the rate with which the mass of that  body increases scales as $\dot{m} = \rho \sigma u$. Similarly, it can be argued that by inertial back-reaction, the force experienced by the object would be \linebreak $F_\mathrm{acc} = - \dot{m} u = -\rho \sigma u^2$~\cite{Macedo:2013qea}. As expected, the magnitude of the effects become larger with the density of the environment and the rate with which particles are introduced to the object, as determined its velocity. This prescription has previously been used to gauge the importance of the effect on dressed inspirals~\cite{Yue:2017iwc}. We will instead only employ it as a stepping stone.

\begin{figure}[t!]
  \centering
  \includegraphics[width=0.92\columnwidth,scale=0.8]{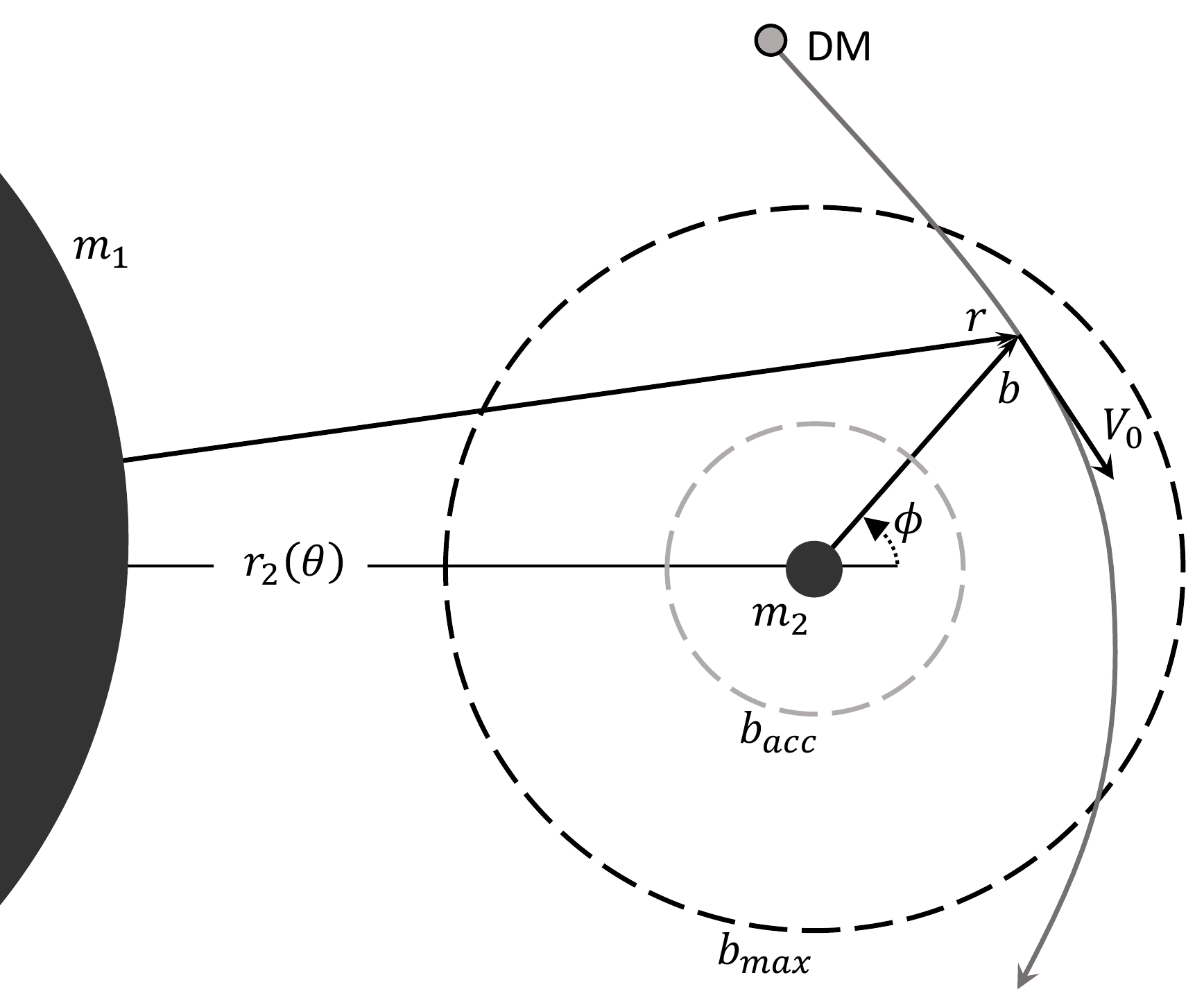}
  \caption{\textbf{Coordinate system for dark matter encounters with the companion.}
  An illustration of a gravitational encounter with a particle in the secondary BH's rest frame. The initial incoming velocity of the DM particle is denoted as $\bm{V_0} = \mathbf{v} -\bm{u}$, and its impact parameter $b$ dictates its future; particles are absorbed within $b_\mathrm{acc}$ or their orbits perturbed up until $b_{max}$. The angle $\theta$ is the binary's true anomaly, while $\phi$ is between the separation $r_2$ and a vector pointing from the central black hole to the location of the impact parameter, $\bm{r} = \bm{r_2} +\bm{b}$.
  }
  \label{fig:scatter}
\end{figure}

For a complete derivation, let us expand the previous picture by instead considering a general particle distribution with vector velocity density $f_{3D}(\mathbf{v}) = \rho\, f(\calE)$ as in \cref{fig:scatter}. Asymptotically and far from the object, a particle will approach with an initial relative velocity $\bm{V_0} = \bm{u} - \mathbf{v}$ and an impact parameter $b$ smaller than the critical value to enable accretion:
\begin{equation}
\label{eq:b_acc}
    b_\mathrm{acc} = \sqrt{\frac{\sigma(V_0)}{\pi}}\,.
\end{equation}
At the end of the encounter, when a particle merges with the test object, conservation of mass dictates that $m \to m +\mu$, while momentum conservation fixes a change in the velocity
\begin{equation}
  \Delta \bm{u} \approx \frac{\mu}{m_{\,\,}} \bm{V_0}\,.
\end{equation}
Ignoring the typical duration of a plunge, at a time interval $\Delta t$ the number of particles the object will absorb from an annulus of radius $b$ with width $\mathrm{d}b$ is given as:
\begin{equation} \label{eq:dN_accretion}
  \Delta N = \frac{1}{\mu} 2\pi b\,\mathrm{d}b V_0 \Delta t \, \rho\, f_{3D}(\mathbf{v}) \,\mathrm{d}^3\mathbf{v}\,.
\end{equation}
Consequently, after the test body has been exposed to all possible particle encounters in this interval, the average mass and absorbed velocity vector would be
\begin{align}
  \langle m \rangle &= \int 2\pi b\,\mathrm{d}b V_0 \Delta t \, \rho\, f_{3D}(\mathbf{v}) \,\mathrm{d}^3\mathbf{v}\,, \\
  \langle \Delta\bm{u} \rangle &= \frac{1}{\mu} \int 2\pi b \,\mathrm{d}b \Delta \bm{u} V_0 \Delta t \, \rho\, f_{3D}(\mathbf{v}) \,\mathrm{d}^3\mathbf{v}\,.
\end{align}
Alternatively, by evaluating the integration over the impact parameters $b$ and re-arranging the time infinitesimals $\Delta t$ we construct the total mass accretion rate $\dot{m}$ and accumulated force of accretion $\bm{F}_\mathrm{acc}$:
\begin{align} \label{eq:accretion_rates}
  \dot{m} = \rho\, &\int \sigma(V_0) V_0 \, f_{3D}(\mathbf{v}) \,\mathrm{d}^3\mathbf{v}\,, \\ \label{eq:accretion_rates_}
  \bm{F}_\mathrm{acc} = \rho\, &\int \sigma(V_0) V_0 \bm{V_0} \, f_{3D}(\mathbf{v}) \,\mathrm{d}^3\mathbf{v}\,.
\end{align}
To avoid the unwieldy expressions above, we may re-define them (recalling the arguments from dimensional analysis above) as:
\begin{equation}\label{eq:accretion_rates2}
\begin{aligned}
  \dot{m} = &\rho\, \sigma(u) u\, \mathcal{C}_m \,, \\
  \mathbf{F}_\mathrm{acc} = -&\rho\, \sigma(u) u^2\, \mathbfcal{C}_\mathrm{acc}\,.
\end{aligned}
\end{equation}
The pre-factors $\mathcal{C}_m$ and $\mathbfcal{C}_\mathrm{acc}$ represent the integrations over the velocity distribution and as such the state of the spike; they are akin to the coefficient $\mathcal{C}_\text{DF}$ appearing in \cref{eq:DF}. While the first factor can only be a scalar, the second is generally a vector indicating the direction of the force.

Note that, in contrast to the aforementioned depiction by dimensional analysis, this force is not simply an inertial drag, but exists alongside the mass accretion rate and indeed, the two quantities are not necessarily proportional to one another ($\mathcal{C}_m \neq |\mathbfcal{C}_\mathrm{acc}|$). Furthermore, while $\dot{m}$ can only be positive, the direction of $\bm{F}_\mathrm{acc}$ is generally sensitive to the specific vector velocity distribution $f_{3D}(\mathbf{v})$ of the halo.\\

\paragraph{\bf Isotropic Spikes.}
For a particle distribution that is isotropic, the
integrations can be considerably reduced to a single evaluation over the magnitude velocity distribution $f(v)$~\cite{Edwards:2019tzf}. Additionally, isotropy constrains the force to act only opposite to the direction of motion of the test body $\bm{u}$, thus $\mathcal{C}_\mathrm{acc}$ becomes a scalar. Thus, the resulting dissipative force behaves like a friction, which we may refer to as ``accretional friction''. It can be written as:
\begin{align} \label{eq:accretion_force}
  \bm{F}_\mathrm{acc} = - \rho\, \sigma(u) u^2\, \mathcal{C}_\mathrm{acc} \bm{\hat{u}}\,.
\end{align}

We now derive the accretion pre-factors by considering the functional form of the accretion cross-section $\sigma$. Utilizing \cref{eq:c_section} we solve for the pre-factors \cref{eq:accretion_rates2} through the integration in \cref{eq:accretion_rates,eq:accretion_rates_}:
\begin{align} \label{eq:coeffs_definition}
  \mathcal{C}_m = &\int \frac{\sigma(V_0) V_0}{\sigma(u) u} \, f_{3D}(\mathbf{v}) \,\mathrm{d}^3\mathbf{v}\,, \\ 
  \mathcal{C}_\mathrm{acc} = &\int 
  \frac{\sigma(V_0) V_0}{\sigma(u) u^2} \left(-\bm{V_0}\cdot\hat{\bm{u}}\right) \, f_{3D}(\mathbf{v}) \,\mathrm{d}^3\mathbf{v}\,.
\end{align}
We perform the integration in spherical coordinates $\mathrm{d}^3\mathbf{v} = v^2 \, \mathrm{d}v\, \mathrm{d}(\cos\gamma) \, \mathrm{d}\phi$, trivially evaluating it over $\phi$, and substituting for $V_0$ through $\cos\gamma = (u^2 +v^2 -V_0^2)/(2uv)$. The final integration remains over the magnitude of the particles' velocity and requires knowledge of the distribution function, $f_{3D}(\mathbf{v}) = f(v)/(4\pi v^2)$ which can be expressed through its moments. We find:
\begin{widetext}
  \begin{align} \label{eq:accretion_ratesproper}
    \mathcal{C}_m &= \xi_{<u} + \frac{\xi^{(2)}_{<u} +\big( 3k +\beta^2 \big) \beta \, \xi^{(-1)}_{>u} +3\beta^2 \xi^{(1)}_{>u}}{3(k +\beta^2)}\,,
    \\
    \mathcal{C}_\mathrm{acc} &= \xi_{<u} + \frac{\big( 10\beta^2 -5k\big) \xi^{(2)}_{<u} -\xi^{(4)}_{<u} +\big( 4\beta^2 +10k\big)\beta^3 \, \xi^{(-1)}_{>u} +20 \beta^3 \xi^{(1)}_{>u}}{15\beta^2(k +\beta^2)}\,,\label{eq:accretion_ratesproper2}
  \end{align}
\end{widetext} 
with $k=r_s/r_\mathrm{acc}$ and $\beta = u/c$.
The coefficients $\xi^{(n)} \equiv \avg{\frac{v^n}{c^n}}$ denote skewed moments of the normalized magnitude velocity distribution, faster ``$>u$", or slower ``$<u$" than the test body's velocity $u$, with $\xi^{(0)} \equiv\xi$. We point out how both of the terms receive a direct contribution from $\xi_{<u}$ which is somewhat analogous to the leading order  dynamical friction's $\mathcal{C}_\mathrm{DF}$~\cite{Dosopoulou_2017}, albeit sub-dominant in this case. As a cross-check, setting $f(v)\propto \delta(v)$, or equivalently sending all particles to $v\ll u$, we recover $\mathcal{C}_m=\mathcal{C}_\mathrm{acc}=1$.

We will use \cref{eq:accretion_ratesproper,eq:accretion_ratesproper2} in \cref{sec:n_bodies} to validate our approach numerically before implementing them in binary evolution in \cref{sec:circ_acc}. However, because of the non-trivial co-dependence of DM distribution function, it is not possible to evaluate these coefficients analytically, expect for in the case of static spikes.

\subsection{Orbital evolution with variable mass}

We now derive the equations of motion describing the binary, including environmental effects and a variable companion mass. 

When the binary's total mass is not conserved, the equations of motion in \cref{eq:da_dtdedt} will be altered. Let us observe what happens to the orbital energy of the system (\cref{eq:energy_1,eq:energy_2}) in a small time-step $\delta t$ where the companion's velocity and mass changes:
\begin{equation}
\label{eq:dE_dt}
  \frac{1}{2}\delta{m}_2 u_2^2 +m_2 \delta{u}_2 u_2 -\frac{Gm_1\delta{m}_2}{r} = E \bigg(\frac{\delta{m}_2}{m_2} -\frac{\delta{a}}{a}\bigg)\,,
\end{equation}
where we assume $u_1$ and $r$ to remain constant during an encounter. After identifying the term $m_2 \delta{u}_2 u_2$ as the work done $\delta{E}$ by forces acting on the companion, the change in the semi-major axis is
\begin{equation}
\label{eq:deltaa_dt}
  \delta{a} = -a \bigg[ \frac{\delta{E}}{E} +\frac{\delta{m}_2}{m} \bigg( \frac{2a}{r} -1 \bigg) \bigg]\,.
\end{equation}

Similarly, to derive the change in eccentricity we start from the orbital angular momentum $\bm{L}$ and obtain,
\begin{equation}
\label{eq:dL_dt}\scalebox{1}{$\displaystyle
\frac{m_1}{m} \bigg( \delta{L} + \frac{\delta{m}_2}{m_2}L \bigg)\! = \frac{L}{2} \bigg( 2\frac{\delta{\mu}}{\mu} +\frac{\delta{a}}{a} +\frac{\delta{m}}{m} -\frac{2\,e\, \delta{e}}{1-e^2} \bigg),$}
\end{equation}
and the infinitesimal is found to be
\begin{equation}
\label{eq:deltae_dt}
\delta{e} = -\frac{1-e^2}{e} \bigg[ \frac{\delta{E}}{2E} + \frac{\delta{L}}{L} +\frac{\delta{m}_2}{m}\bigg( \frac{a}{r} -1 \bigg) \bigg]\,.
\end{equation}

The effects of the non-conservation of mass are generally suppressed by the binary's small mass ratio $q$ because of the terms $\dot{m}_2/m$. However, the force of accretion $F_\mathrm{acc}$ is not suppressed in this way, and changes to the companion's mass can alter the gravitational wave emission and dynamical friction of the system.

To derive a differential equation from the infinitesimals of \cref{eq:deltaa_dt,eq:deltae_dt}, we integrate the cumulative variation for the duration of a single orbit. Hence, for example, the total change in the semi-major axis is $\Delta a = \int \delta a \, \mathrm{d}t/T$, and due to the secular evolution of the system we may also write $\dot{a} \approx \Delta a/T$. To implement this, instead of integrating in time, we substitute to the full circle of the true anomaly utilizing $\dot{\theta} = \sqrt{Gma(1-e^2)}/r^2$~\cite{Maggiore:2007ulw} and \cref{eq:separation,eq:period}, or:
\begin{equation}
  \label{eq:averageEnergy}
  \int \displaylimits_0^{T} {...\,\,} \frac{\mathrm{d}t}{T} = (1 -e^2)^{3/2} \int \displaylimits_0^{2\pi} {...\,\, (1 +e\cos\theta)^{-2}} \frac{\mathrm{d}\theta}{2\pi} \,.
\end{equation}
Thus, the system of equations becomes
\begin{align}
\label{eq:da_dt}
  \dot{a} &= -a \bigg[ \frac{\dot{E}}{E} +\frac{2\dot{\bar{m}}_2 - \dot{m}_2}{m} \bigg]\,,\\
  \label{eq:de_dt}
  \dot{e} &= -\frac{1-e^2}{e} \bigg[ \frac{\dot{E}}{2E} + \frac{\dot{L}}{L} +\frac{\dot{\bar{m}}_2 - \dot{m}_2}{m} \bigg]\,,
\end{align}
where each of the derivatives on the right-hand side are calculated through \cref{eq:averageEnergy}, and the term $\dot{\bar{m}}_2$ is the same but for the orbit-weighted product $\dot{m}_2 \,r/a$. The latter term appears due to the action of accretion on the secondary and is, thus, not present in previous work \cite{Munoz_2019,DOrazio_2021,10.1093/mnras/stad1131}.

\subsection{Mass accretion feedback on the spike}
\label{sec:massdepletion}

We will now study how accretion by the secondary depletes the distribution of a dark matter halo. In \citet{Kavanagh:2020cfn}, we introduced a feedback model that alters the distribution function $f(\calE)$ of the spike to account for the deflection of particles (dynamical friction) with the companion. Below, we expand on this methodology to also account for the removal of particles after falling inside the companion's horizon.

Our aim is to derive a differential equation $\dot{f}(\calE)$ which describes this mode of particle depletion. We focus on the change in the spike distribution over a single orbit of the companion, such that due to the secular evolution of the system, we can write: 
\begin{equation}
  \dot{f}(\calE) \approx\nolinebreak \frac{\Delta f(\calE)}{T}  \,.
\end{equation} 
Then, to derive a prescription for $\Delta f$, we find the number of particles $\Delta N_\mathrm{acc}(\calE) =\nolinebreak g(\calE)\,\Delta f(\calE)/\mu$ falling inside the companion during this orbit. Here, $g(\calE)$ is the density of states (DoS) of the DM spike. We can also evaluate this fraction $\Delta N_\mathrm{acc}(\calE)=\nolinebreak p_\mathrm{acc}(\calE) N(\calE)$ using the probability $p_\mathrm{acc}(\calE)$ that an orbit falls inside the horizon.

We can write the accretion probability as the fraction of orbits with energy $\calE$ that satisfy the capture condition $b \leq b_\mathrm{acc}$:
\begin{equation}
  \label{eq:bayesProb}
  p_\mathrm{acc}(\calE) = \frac{g'(\calE | b \leq b_\mathrm{acc})}{g(\calE)}\,,
\end{equation}
where $g'(\calE | b \leq b_\mathrm{acc})$ is a `restricted' density of states, subject to the restriction that $b \leq b_\mathrm{acc}$.

The total density of states is defined as an integral over phase space of all configurations which can occupy a specific energy $\calE$~\cite{BinneyAndTremaine}:
\begin{equation}
  \label{eq:DoS0}
  g(\calE) \equiv \iint \delta(\calE -\Psi(r)+v^2/2) \mathrm{d}^3\bm{r} \, \mathrm{d}^3\mathbf{v}\,.
\end{equation}
For the dark dress, whose gravitational potential $\Psi(r)$ is dominated by the central black hole within its size, we find~\cite{Kavanagh:2020cfn}:
\begin{equation}
  g(\calE) = \sqrt{2} \pi^3 G^3 m_1^3 \calE^{-5/2} \,,
\end{equation}
which only depends on the mass $m_1$ and not on the properties of the spike.

For the restricted DoS we follow a similar integration but include also a Heaviside step function $H(b-b_\mathrm{acc})$, to implement the capture condition:
\begin{align} \label{eq:dos2}
  g' &= \! \iint \delta\big(\calE -\Psi(r)+v^2/2\big) H\big(b-b_\mathrm{acc}\big) \mathrm{d}^3\bm{r} \, \mathrm{d}^3\mathbf{v}\,.
\end{align}
The Heaviside function restricts the spatial volume to that of a deformed torus whose major radius $r_2(\theta)$ follows the trajectory of the companion and whose minor radius is the accretion impact parameter $b_\mathrm{acc}$. In \cref{app:deformed_torus}, we derive the volume element of the deformed torus $d^3\textbf{r}=\nolinebreak b(r_2+b\cos\phi)\, \,\mathrm{d}b\, \mathrm{d}\theta\, \mathrm{d}\phi$, in order to perform the integral over $b$ and $\phi$. While the details of the derivation are described in \cref{app:accretion_probability}, we eventually construct the probability of accretion as:
\begin{equation} \label{eq:accretion_probability}
    p_\mathrm{acc}(\calE) = \frac{4\pi^2}{g(\calE)} \zeta r_\mathrm{acc}^2 \int r_2 \bigg(v_* +\frac{r_s}{r_\mathrm{acc}} \frac{c^2}{2u} \ln \frac{u+v_*}{|u-v_*|} \bigg) \, \mathrm{d}\theta\,,
\end{equation}
where $v_*(\calE,r_2) = \sqrt{2\left( \Psi(r_2) -\calE \right)}$ is the velocity of a particle at a radius $r_2$ with energy per unit mass $\calE$. Finally, by extension the feedback model will be
\begin{equation} \label{eq:accretion_feedback}
    \dot{f}_\mathrm{acc}(\calE) = -p_\mathrm{acc}(\calE) f(\calE)/T\,.
\end{equation}
The final integration over the true anomaly $\theta$ in \cref{eq:accretion_probability} captures depletion from eccentric binary inspirals and is thus vital for the description of eccentricity evolution in dynamic dark matter distributions. Due to its non-trivial form, we perform this integral numerically.

\subsection{Mass conservation}

Before validating our formalism against $N$-body simulations in \cref{sec:n_bodies}, we may first perform an analytical cross check. Mass conservation in the system will dictate that the amount of mass added to the black hole companion $\Delta m_2$ should be the same as that removed from the spike $\Delta m_\mathrm{sp}$ through the accretion feedback model of \cref{eq:accretion_feedback}.

During a single cycle of a quasi-circular inspiral, the companion's mass will roughly increase with \cref{eq:accretion_rates2}~as,
\begin{equation} \label{eq:mass_from_rate}
    \Delta{m}_2 = \rho \, \sigma \, u \, T \, \mathcal{C}_m \,,
\end{equation}
where $\rho$ is the DM density at the orbital radius, $T$ is the orbital period, and $\mathcal{C}_m$ is the mass accretion pre-factor of \cref{eq:accretion_ratesproper}, sensitive to the spike's distribution function. Similarly, the mass removed from the spike $\Delta m(\calE)$ is found at each specific energy $\calE$, as $\Delta m(\calE)=\nolinebreak \mu \Delta N_\mathrm{acc}(\calE)=\nolinebreak g(\calE) f(\calE) p_\mathrm{acc}(\calE)$. Integrating over all energies gives the total mass $\Delta m_\mathrm{sp}$,
\begin{equation} \label{eq:spike_mass0}
  \Delta m_\mathrm{sp} = \int_0^{\Psi(r_2)} \Delta m(\calE)\, \mathrm{d}\calE\,.
\end{equation}
Combining \cref{eq:accretion_probability,eq:spike_mass0} we derive
\begin{align} 
\begin{split} 
\label{eq:spike_mass}
  \Delta m_\mathrm{sp} &= 8 \pi^3 \zeta r^2_\mathrm{acc} r_2 \times\\
  &\qquad\int_0^{\Psi} \left( v_* +\frac{r_s}{r_\mathrm{acc}} \frac{c^2}{2u} \ln\frac{u+v_*}{|u-v_*|} \right) f(\calE) \,\mathrm{d}\calE\,.
\end{split}
\end{align}
Given that for a single circular orbit $2 \pi r_2 = u T$, we can see how \cref{eq:spike_mass} takes the form:
\begin{equation} \label{eq:spike_mass2}
    \Delta{m}_{sp} = \rho \, \sigma \, u \, T \, \mathcal{C}'_m \,.
\end{equation}
Comparing with \cref{eq:mass_from_rate}, we see that $\Delta{m}_{sp} \neq \Delta{m}_2$, as long as $\mathcal{C}_{m} \neq \mathcal{C}'_{m}$. While this is generally the case, the two coefficients are similar, roughly within $10-20\%$ (see \cref{fig:accretion_force_test}). We attribute this inconsistency to the analytical approximation of setting $v_*(\calE, r) \to v_*(\calE, r_2)$, taken in evaluating \cref{eq:dos2}. We note also that, as we will see in \cref{sec:n_bodies}, the feedback is in line with our numerical validations. Additionally, we can check that in the small particle velocity limit $v_* \ll u$, \cref{eq:spike_mass} becomes,
\begin{align} \label{eq:spike_mass_laminar}
  \Delta m_\mathrm{sp} &= 8 \pi^3 \zeta r^2_\mathrm{acc} r_2 \left( 1 +\frac{r_s}{r_\mathrm{acc}} \frac{c^2}{u^2} \right) \int_0^{\Psi} v_* f(\calE) \,\mathrm{d}\calE \nonumber \\
  & = \rho \sigma u T \,,
\end{align}
which is the expected result for the steady laminar flow toy model, lacking knowledge of all possible particle encounters, and equivalently all factors $\mathcal{C}_m$, $\mathcal{C}'_m$ reduce to one.

\section{Dark matter accretion in N-body simulations}
\label{sec:n_bodies}

In order to validate the analytic formalism which has been developed so far, we perform $N$-body simulations of binaries embedded in DM spikes using the publicly available \texttt{NbodyIMRI} code~\cite{NbodyIMRI}. This code is introduced in the companion paper (Ref.~\cite{BetterSpikesII}), where we provide full details, including the numerical checks which have been performed on the code. Here, we provide a brief summary.

\texttt{NbodyIMRI} uses a 4th-order leapfrog integration scheme with fixed timesteps to follow the dynamics of the two BHs ($m_1$ and $m_2$) and $N_\mathrm{DM}$ DM (pseudo-)particles. The Newtonian force between the two BHs and between each BH and the DM particles is calculated exactly. However, the pair-wise forces between DM particles are neglected, because the contribution of the DM spike to the gravitational potential should be negligible at distances sufficiently close to the BH binary. With this approximation, it is possible to perform direct simulations of large numbers of DM particles $N_\mathrm{DM} \sim \mathcal{O}(10^6)$ with small timesteps, in order to accurately resolve the dynamics of the DM particles close to the BHs. 

\texttt{NbodyIMRI} uses a global time-step, set by choosing $N_\mathrm{step} = 10^4$ time-steps per orbit of the binary. With fixed time-steps, very close encounters between DM particles and the lighter black hole cannot be accurately resolved. We therefore introduce a softening which cuts off the gravitational force exerted by $m_2$ at distances $r < r_\mathrm{soft}$. We set the softening length by requiring that the time for DM particles to cross a distance $r \sim r_\mathrm{soft}$ close to the secondary BH should always be longer than one time-step. With this requirement, we set  $r_\mathrm{soft} = 10^{-3} \,a_i$ for an initial semi-major axis of $a_i$. See Paper II~\cite{BetterSpikesII} for a full discussion of the choice of softening lengths.

For stellar-mass BHs, the accretion radius $r_\mathrm{acc} \sim r_s \sim \mathrm{km}$. The typical length-scales of the simulation will be comparable to the BH separation $r_2 \sim 100 \,r_{\mathrm{ISCO}}$, where $r_{\mathrm{ISCO}} = 6 G m_1/c^2$ is the innermost stable circular orbit (ISCO) of the primary BH. For $m_1 = 1000 \,M_\odot$, we have $r_2 \sim 10^5\,\mathrm{km}$. Assuming $N_\mathrm{DM} = 10^6$ particles distributed uniformly in this volume, we find a mean inter-particle spacing of $\sim 10^3\,\mathrm{km}$, compared to the km-scale accretion radius. Thus, even with a large number of particles in the simulation, the probability of accretion for a realistic BH would be negligible. 

In order to verify the dynamics of the accretion process then, we use an artificially large accretion radius $r_\mathrm{acc} \gg r_s$ ($k \ll 1$) for the secondary BH.\footnote{We neglect accretion onto the primary BH.} This will allow us to verify the feedback formalism described in \cref{sec:massdepletion}, as well as testing the scaling of the accretion force with $r_\mathrm{acc}$, which can then be extrapolated down to the smaller radii relevant for real BHs. In practice, at each timestep of the simulation, we check whether a DM particle lies within a distance $r < r_\mathrm{acc}$ of the secondary BH. If it does, we remove the particle from the simulation, add its mass to the secondary BH and update the BH velocity $\bm{u}_2$ such that the momentum of the system is conserved. With the accretion of a DM pseudo-particle of mass $\tilde{m}_\mathrm{DM}$ with a velocity $\mathbf{v}$, the BH velocity is updated to:
\begin{equation}
    \bm{u}_2^\prime = \frac{m_2 \bm{u}_2 + \tilde{m}_\mathrm{DM} \mathbf{v}}{m_2 + \tilde{m}_\mathrm{DM}}\,.
\end{equation}

\subsection{Orbit backreaction and mass accretion rate}

We simulate a benchmark system with BH masses $(m_1, m_2) = (1000, 1)\,M_\odot$ on a circular orbit with initial semi-major axis $a_i = 100\,R_{\mathrm{ISCO}}$ for a range of values of $r_\mathrm{acc}$. For each value of $r_\mathrm{acc}$, we generate 16 realisations of a DM spike with $N_\mathrm{DM} = 10^6$ particles. We evolve each realisation of the binary-and-spike system for 1 orbit, such that the results should be unaffected by feedback effects which may accumulate over many orbits.    By measuring the change in semi-major axis $\Delta a/a$ during the simulation, we can estimate the change in the energy of the binary due to environmental effects, following \cref{eq:da_dt}. We note that the change in semi-major axis due to mass accretion $\Delta m_2/m$ is negligible in this case.

\begin{figure}[tb]
    \centering
    \includegraphics[width=0.95\columnwidth]{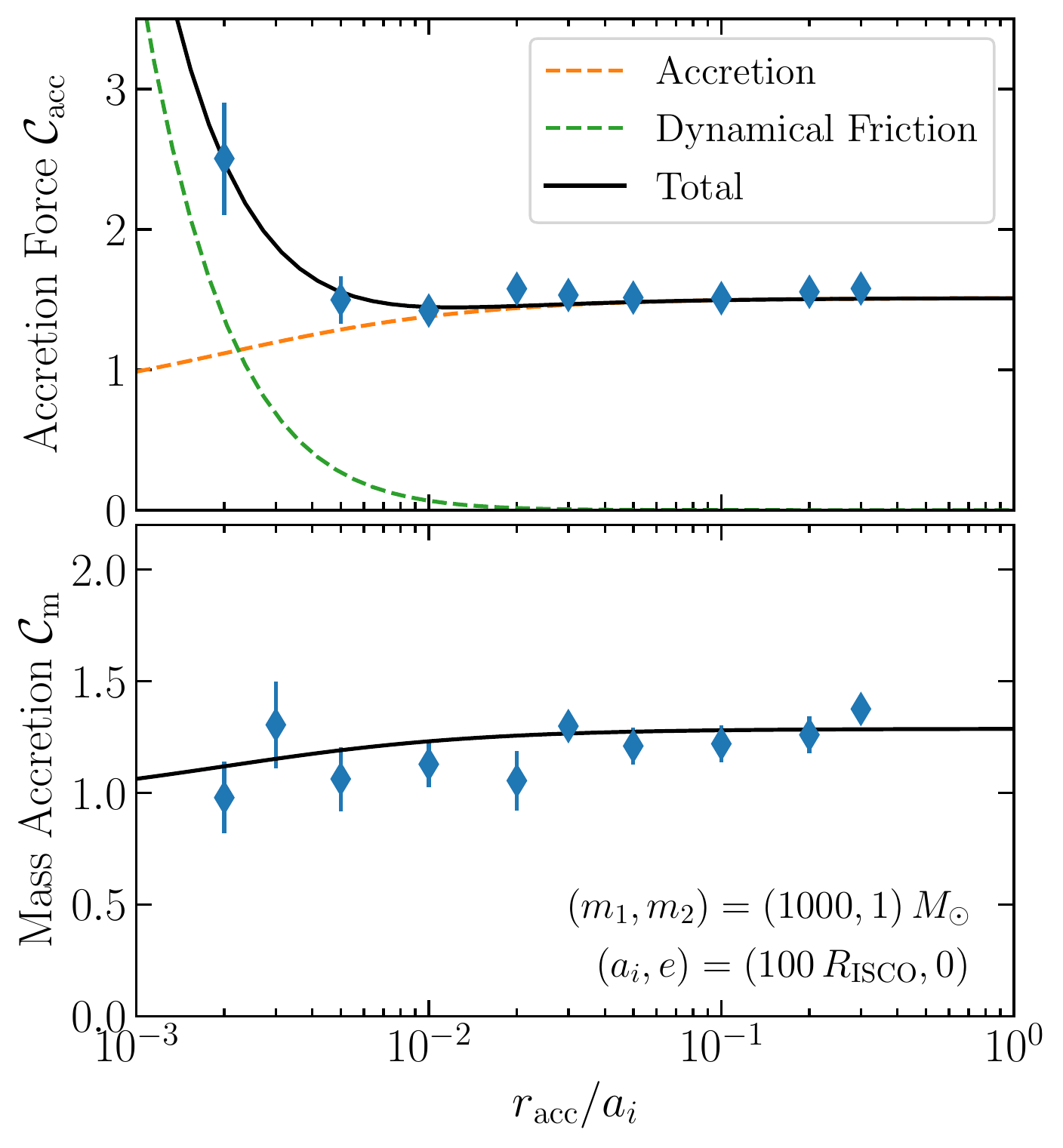}
    \caption{\textbf{Coefficients for the accretion force $\mathcal{C}_\mathrm{acc}$ and mass accretion rate $\mathcal{C}_\mathrm{m}$ as a function of accretion radius $r_\mathrm{acc}$.} The dimensionless accretion coefficients are defined in \cref{eq:accretion_rates2} and the surrounding text. Data points show the accretion coefficients estimated from $\dot{a}/a$ measured in simulations, while solid lines show the values calculated directly using \cref{eq:coeffs_definition}. In the top panel, the dashed green line shows the spurious contribution of dynamical friction to the accretion coefficient inferred from $\dot{a}/a$. For comparison, in the bottom panel, we also show the value of $\mathcal{C}_m'$ in \cref{eq:spike_mass2} as a dotted line.}
    \label{fig:accretion_force_test}
\end{figure}

The results are shown in \cref{fig:accretion_force_test}. In the upper panel, we show the accretion coefficient inferred from the simulations, assuming accretion is the only environmental effect acting on the binary (blue points). We also compare these results to the theoretical predictions of \cref{eq:accretion_ratesproper,eq:accretion_ratesproper2}. We find that for small values of $r_\mathrm{acc}$, the inferred value of $C_\mathrm{acc}$ rises compared to the predicted value. This is because we have estimated $C_\mathrm{acc}$ from the rate of inspiral of the binary. However, there are two physical effects in the simulation which can lead to the orbital decay of the binary: accretion and dynamical friction. The latter arises from DM particles which scatter gravitationally with small impact parameters (without being accreted) and this effect begins to dominate at small $r_\mathrm{acc}$. In \cref{fig:accretion_force_test}, we have estimated the value of $C_\mathrm{acc}$ assuming that accretion is the only environmental effect at play, which leads to a spurious contribution to $C_\mathrm{acc}$ coming from dynamical friction. Including also this contribution (shown as a dashed green line), we find good agreement with the simulation results.

 By measuring the change in mass of the secondary during the simulation, we can also infer the mass accretion coefficient $\mathcal{C}_\mathrm{m}$, as shown in the lower panel of \cref{fig:accretion_force_test}. We see that there is good agreement between the results of the simulations and the theoretical values of $\mathcal{C}_\mathrm{acc}$ and $\mathcal{C}_\mathrm{m}$. This includes a slight decrease at low values, as gravitational focusing becomes more important. This implies that the formalism we have developed for these forces can be extrapolated to smaller scales, relevant for real BHs.

\begin{figure}[tb]
    \hspace*{-0.8cm}
    \includegraphics[width=0.95\columnwidth]{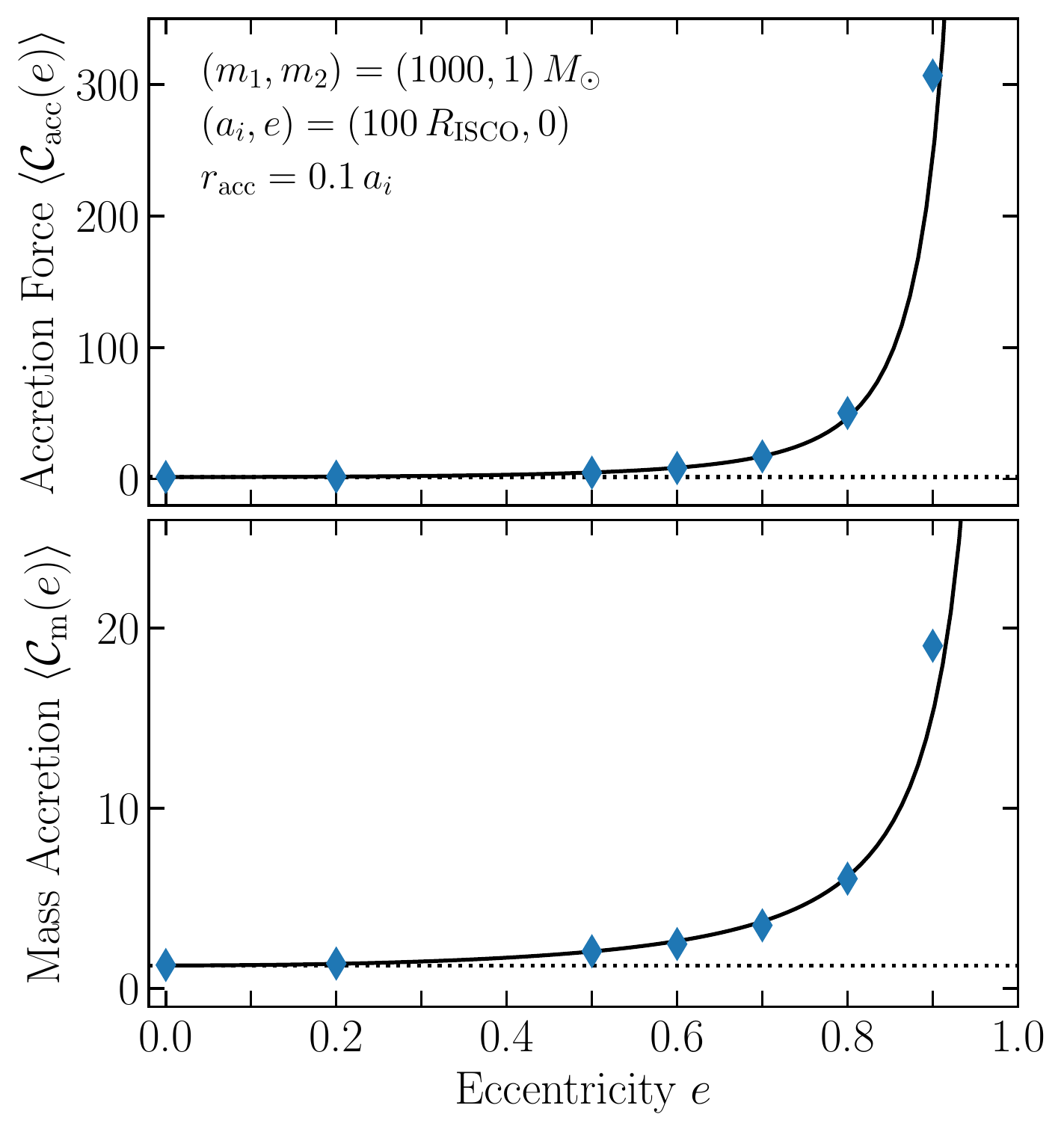}
    \caption{\textbf{Orbit-averaged accretion coefficients $\left\langle \mathcal{C}_\mathrm{acc}(e) \right\rangle$ and $\left\langle \mathcal{C}_\mathrm{m}(e) \right\rangle$ as a function of orbital eccentricity $e$.} Data points show the accretion coefficients estimated from $\dot{a}/a$ measured in simulations, while the solid lines show the values calculated by orbit-averaging of the coefficients defined in \cref{eq:coeffs_definition}. We fix the accretion radius to $r_\mathrm{acc} = 0.1\,a_i$.}
    \label{fig:accretion_force_test_ecc}
\end{figure}

In \cref{fig:accretion_force_test_ecc}, we show the orbit-averaged accretion coefficients $\left\langle \mathcal{C}_\mathrm{acc}(e) \right\rangle$ and $\left\langle \mathcal{C}_\mathrm{m}(e) \right\rangle$ for orbits of eccentricity $e$. These are defined as the values of $\mathcal{C}_\mathrm{acc}$ and $\mathcal{C}_\mathrm{m}$ for a circular orbit which would give the same average accretion force and mass accretion rate as for the eccentric orbit under consideration. By definition, then $\left\langle \mathcal{C}(0) \right\rangle \equiv \mathcal{C}$. 

The numerical values of the accretion coefficients plotted  as blue points in \cref{fig:accretion_force_test_ecc} show a rapid rise at large eccentricity. This is because at large eccentricity the pericentre of the orbit moves to smaller radii, where the DM density is large and the orbital velocity is high, enhancing the accretion rate relative to the circular case. The solid black lines are obtained using the orbit-averaging formalism in \cref{eq:averageEnergy} and show good agreement with the simulated results, even at large eccentricity.

\subsection{Spike depletion}

We now turn to testing the evolution of the dark matter profile itself. To this end, we perform a similar set of $N$-body simulations to numerically recover the profile after a small number of orbits, 
such that the orbital elements of the companion do not change significantly during the simulation.
To obtain a theoretical prediction for the depletion, we can solve \cref{eq:accretion_feedback} trivially to obtain the distribution function at a later time $t$,
\begin{equation} \label{eq:df_accretion}
  f(\calE, t) = f(\calE, 0) \exp\bigg(-\frac{p_\mathrm{acc}(\calE)}{T} t\bigg)\,,
\end{equation}
where $f(\calE, 0)$ is the distribution at the start of the simulation $t=0$. By integrating the distribution function in \cref{eq:df_accretion} over velocities, we can thus compute the depleted density profile $\rho(r, t)$. 

\begin{figure}[tb]
    \hspace*{-0.3cm}
    \includegraphics[width=\columnwidth]{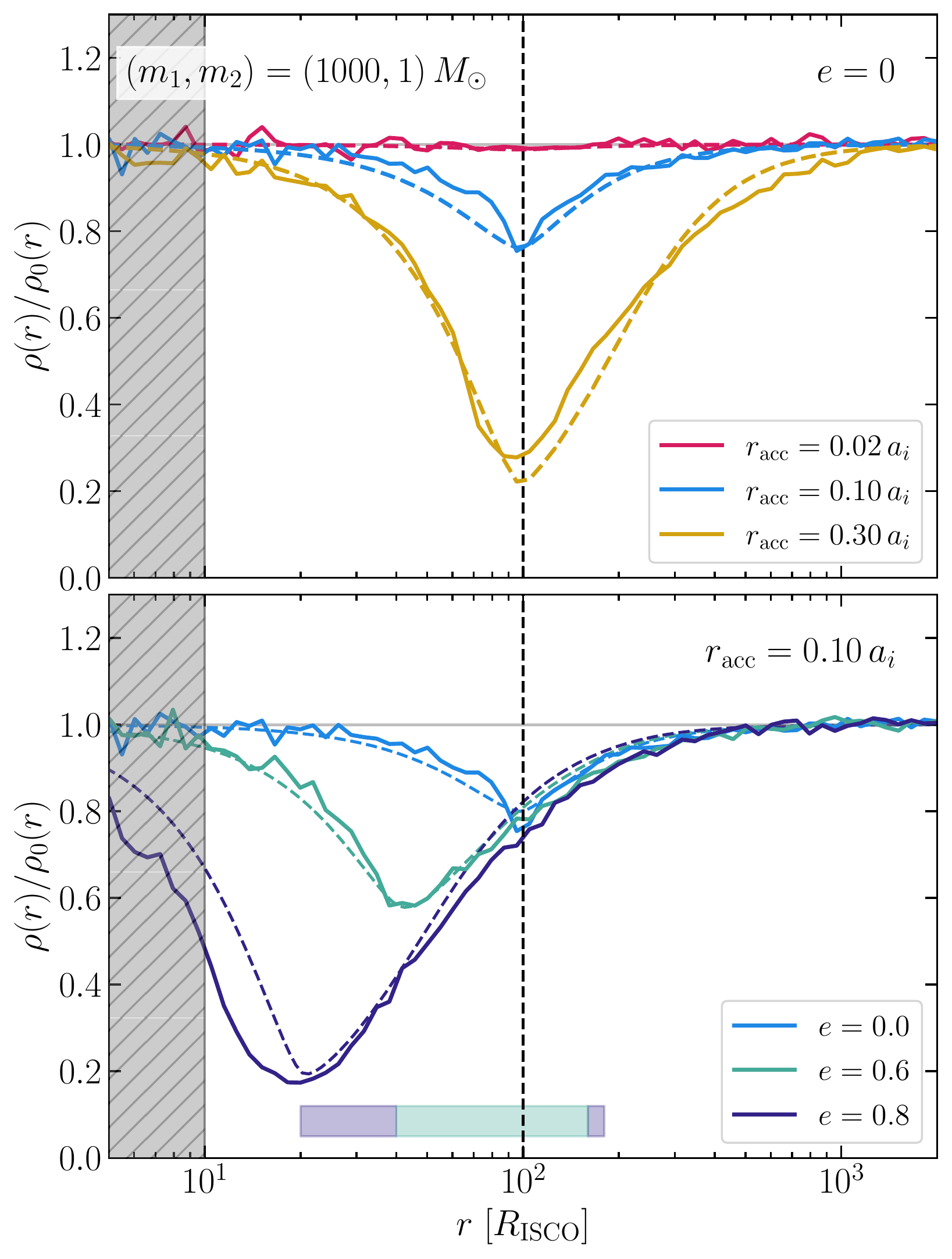}
    \caption{\textbf{Feedback due to accretion as a function of accretion radius and eccentricity.} In the \textbf{upper panel}, we show the ratio of the final to initial DM density profiles as a function of radius after 25 orbits. Solid lines show the results estimated from the \texttt{NbodyIMRI} simulations, while dashed lines are derived from the feedback model of \cref{eq:df_accretion} and \cref{sec:massdepletion}. Different colours correspond to different accretion radii $r_\mathrm{acc}$. In the \textbf{lower panel}, we fix the accretion radius to $r_\mathrm{acc} = 0.1\,a_i$ and show results for different eccentricities $e = \left\{0.0, 0.6, 0.8\right\}$. The coloured bars at the bottom of the panel show the range of radii traversed by the secondary BH over the orbits with eccentricities $e = 0.6$ and $e = 0.8$. The grey shading shows the region where the interactions between the DM particles and the central BH are softened.}
    \label{fig:feedback_test}
\end{figure}

In \cref{fig:feedback_test}, we show a validation of our semi-analytic feedback prescription with numerical results after 25 orbits. In the top panel, we show the depletion of the DM spike for a circular orbit, for different values of the accretion radius $r_\mathrm{acc}$. This depletion shows good agreement (at the level of $5-10\%$) with the semi-analytic formalism described above (coloured dashed lines). Note also that the spike is depleted not only within the accretion radius (which in this case ranges from 2\%-30\% of the orbital radius) but instead there is substantial depletion out to radii an order of magnitude smaller and larger than the orbital radius (vertical dashed line). This occurs because all orbits which pass through the torus where accretion is effective will be depleted, leading to depletion also at smaller and larger radii. 

In the bottom panel of \cref{fig:feedback_test}, we fix the accretion radius to $r_\mathrm{acc} = 0.1\,a_i$ and show results for different eccentricities. The two coloured bars at the bottom of the panel show the range of orbital radii for the two orbits with eccentricities $e = 0.6$ and $e = 0.8$. Again, there is good agreement with the semi-analytic feedback accretion formalism (coloured dashed lines). The only substantial discrepancy appears for the most eccentric orbit ($e = 0.8$) at small radii, where the pericentre $0.2 \,a_i$ becomes comparable to the accretion radius and the softening length of interactions with the central BH, both equal to $0.1 \,a_i$. The largest depletion occurs close to the pericentre, where the orbital velocity is large, in agreement with our interpretation of \cref{fig:accretion_force_test_ecc}, where the accretion force and mass accretion rate are dominated by the inner parts of the orbit.

\begin{figure}[tb]
    \hspace*{-0.4cm}
    \includegraphics[width=\columnwidth]{Figures/Density_evolution_joint.pdf}
    \caption{\textbf{Simulated and predicted DM depletion and orbital decay with time.} The upper panel shows the ratio of the final to initial DM density at the orbital radius of the binary (marked by the vertical dashed line in the upper panel of \cref{fig:feedback_test}). Crosses ($\bm{+}$) show the simulation results, which initially agree with the theoretical feedback formalism (colored dashed lines). The lower panel shows the corresponding fractional change in semi-major axis in the simulations (solid lines) compared to the prediction from the feedback formalism (dashed). For large numbers of orbits, the simulated and theoretical results deviate as the assumptions of isotropy and spherical symmetry are eventually broken.}
    \label{fig:feedback_test_orbits}
\end{figure}

The feedback results so far have focused on a relatively small, fixed number of orbits. We now explore the evolution of the density profile over longer times. For each of the systems in the upper panel of \cref{fig:feedback_test}, we continued evolving the system up to 250 orbits. The results are shown in the top panel of \cref{fig:feedback_test_orbits}, where the crosses show the ratio of the initial and final DM density at the orbital radius $r = a_i$, as measured in the simulations. The dashed lines show the corresponding density calculated using the semi-analytic prescription of \cref{eq:df_accretion}. Initially, this prescription provides a good fit to the simulation results. However, after some number of orbits, the DM density measured in simulations begins to flatten, while the theoretical model predicts an ever decreasing density.

\begin{figure}[tb]
  \begin{center}
      \includegraphics[width=\columnwidth]{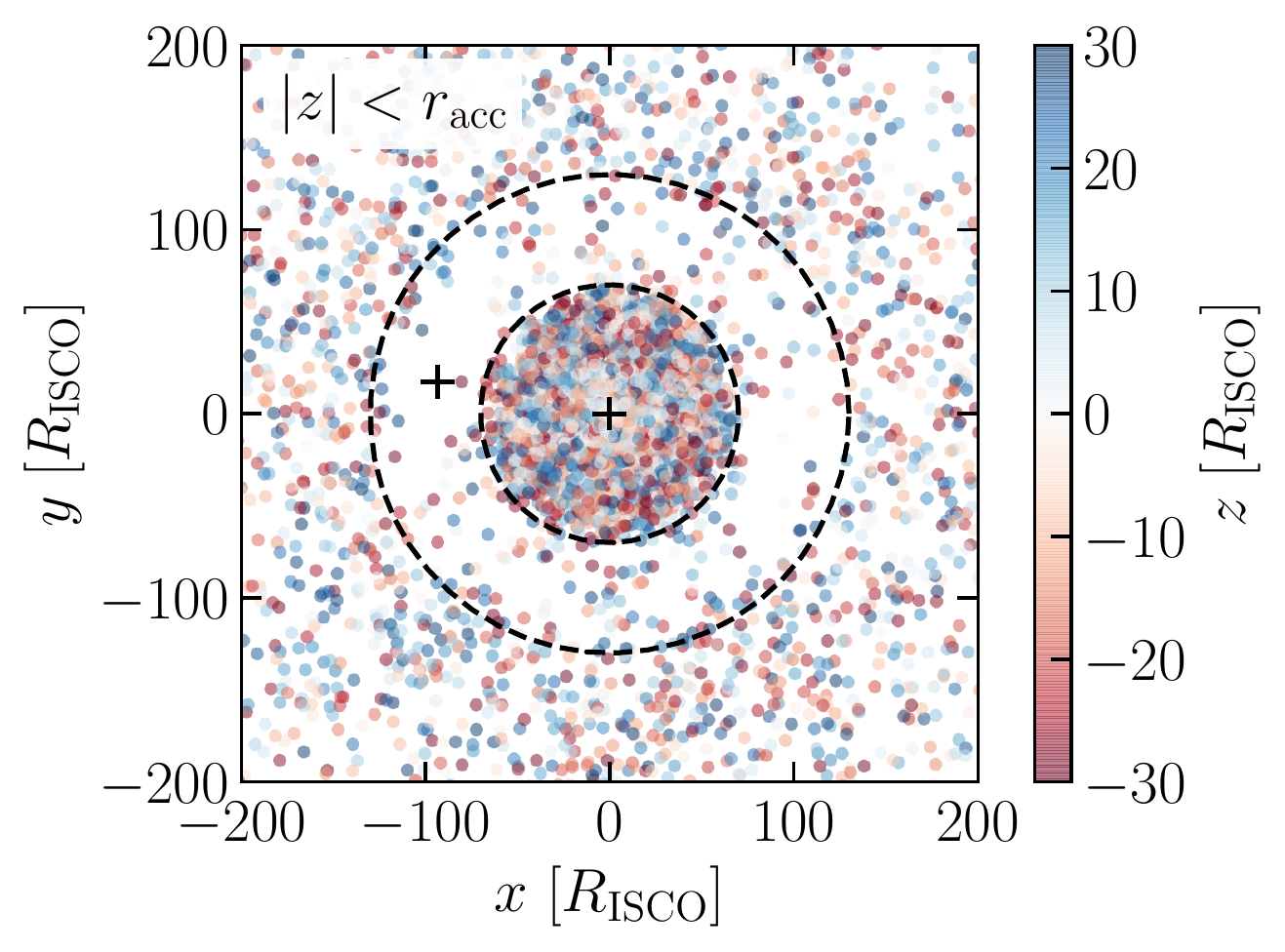}
      \hspace*{-1cm}
      \includegraphics[width=0.8\columnwidth]{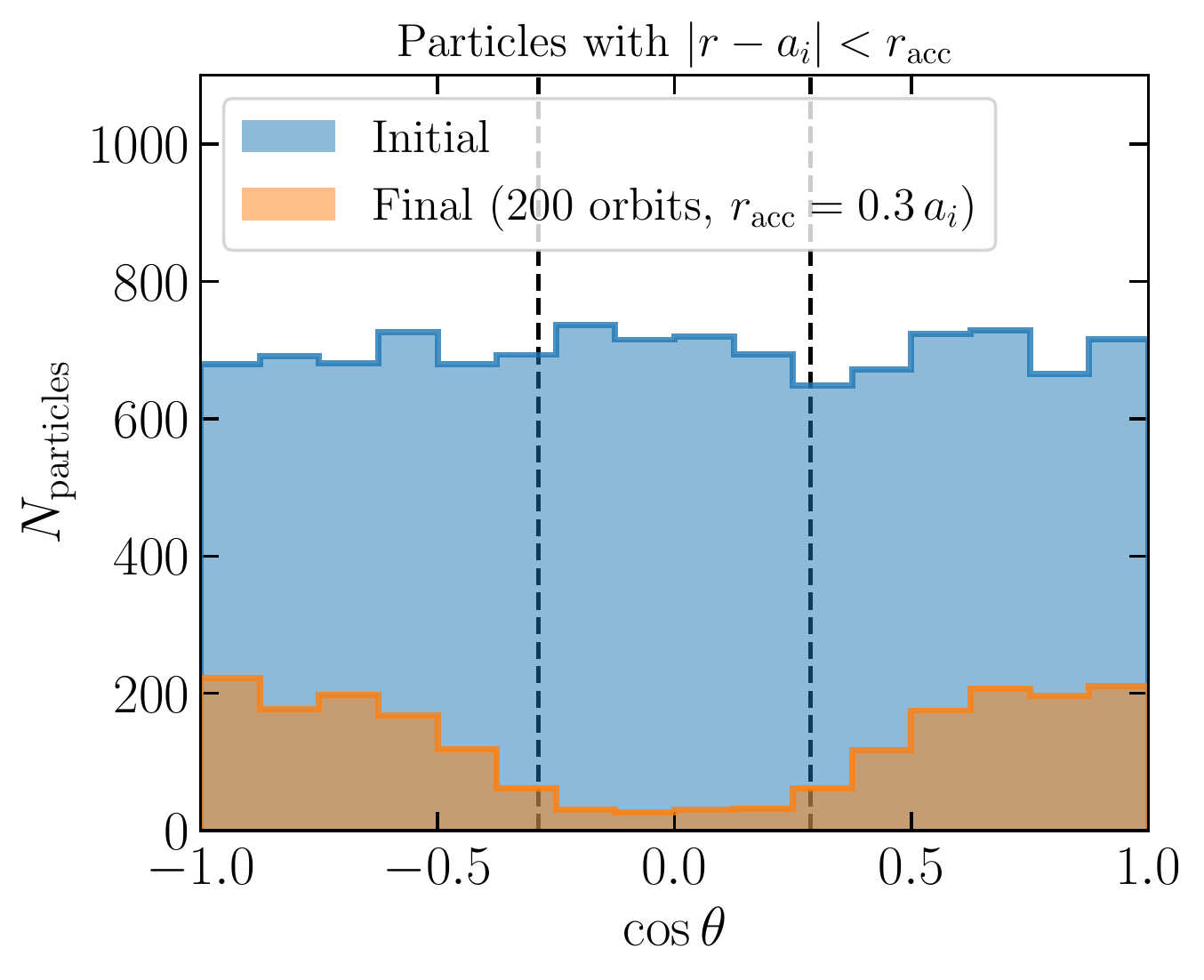}
      \caption{\textbf{Distribution of simulated DM particles after 200 orbits.} The upper panel shows the distribution of particles in the $x$-$y$ plane (the plane of the orbit), limiting to particles with $|z| < r_\mathrm{acc}$. The black crosses show the positions of the two BHs while the dashed region marks the region from which particles may be accreted over a single orbit (i.e.\ the region with a perpedicular distance from the orbital path of less than $r_\mathrm{acc}$). The lower panel shows the distribution of the polar angle $\cos\theta$ of DM particles (where $\cos\theta = 0$ lies in the plane of the orbit), limiting to particles within a spherical shell of width $2\,r_\mathrm{acc}$ around the orbital radius. This snapshot shows results from the simulation with $r_\mathrm{acc} = 0.3\,a_i$ (yellow results in \cref{fig:feedback_test_orbits}).}
      \label{fig:feedback_test_distribution}
  \end{center}
\end{figure}

This deviation between the model and the simulation arises because the assumption of isotropy and spherical symmetry is eventually broken. This can be seen clearly in \cref{fig:feedback_test_distribution}, which shows the distribution of DM particles after 200 orbits for the case of $r_\mathrm{acc} = 0.3\,a_i$ (yellow results in \cref{fig:feedback_test_orbits}). In the upper panel of \cref{fig:feedback_test_distribution}, we show the distribution of particles in the $x$-$y$ plane (the plane of the orbit). This shows that after $\sim 200$ orbits, the orbiting object has largely cleared the region accessible to it via accretion, making further depletion difficult. In the lower panel of \cref{fig:feedback_test_distribution}, we show the distribution of the polar angle $\cos\theta$ of DM particles (where $\cos\theta = 0$ lies in the plane of the orbit). Here, we see that an initially uniform distribution of particles has been substantially depleted, leaving very few particles in the plane of the orbit. 

This snapshot highlights that the accretion process preferentially depletes particles from the torus swept out by the orbit of the secondary BH, with inner cross section set by the accretion radius $r_\mathrm{acc}$. Eventually the number of particles in the torus becomes very small, and nothing more can be accreted. The formalism described by \cref{eq:df_accretion} assumes that the un-accreted DM particles lie on orbits which remain spherically symmetric and therefore there will always be particles crossing this `accretion torus', allowing for continuing depletion.

The lower panel of \cref{fig:feedback_test_orbits} shows the fractional change in semi-major axis of the simulated binaries due to accretion by the secondary (assuming a spike normalisation of $\rho_6 = 9.1 \times 10^{14} \,M_\odot/\mathrm{pc}^3$). As for the DM density, the simulated orbital decay (solid lines) initially follows that expected from an analytic estimate including the feedback formalism described above (dashed lines). In the simulations, the orbital decay begins to stall at roughly the same time that the DM density flattens off in the upper panel. Again, this is due to the asymmetry which develops in the system: depletion stops because no further particles can be accreted, which also means that the inspiral (driven by accretion) stalls.

The number of orbits $N_\mathrm{crit}$ before this asymmetry becomes relevant can be estimated by comparing the volume of the accretion torus $V_\mathrm{torus} = 2 \pi^2 r \,r_\mathrm{acc}^2$ with the volume of a shell of thickness $2\,r_\mathrm{acc}$ at a radius $r$, $V_\mathrm{shell} = 8 \pi r^2 \,r_\mathrm{acc}$. The ratio of these two quantities is an estimate of the probability $p_\mathrm{cap}$ that a given particle in the shell will be captured in a single orbit (i.e.\ if it lies within the accretion torus). The number of orbits before the accretion torus is fully depleted (and further accretion is not possible) therefore goes as $\sim 1/p_\mathrm{cap}$, meaning that $N_\mathrm{crit} \sim V_\mathrm{shell}/V_\mathrm{torus} \sim (4/\pi)(r/r_\mathrm{acc})$. For the examples in \cref{fig:feedback_test_orbits}, this would corresponds to $N_\mathrm{crit} = \left\{ 64, 13, 4\right\}$ for $r_\mathrm{acc} = \left\{0.02, 0.10, 0.30\right\} r$ respectively, roughly matching the number of orbits at which the simulations begin to show deviations from the prediction in each case. 

For the realistic case of accretion by a Schwarzschild black hole, we can set $r_\mathrm{acc} \approx 2 r_s \approx 4 G m_2/c^2$. We note that the secondary BH will not be able to efficiently deplete the accretion torus if it inspirals by more than a distance $ 2 r_s$ during $N_\mathrm{crit}$ orbits. The number of orbits required for the secondary BH to inspiral due to GW emission by a distance $2 r_s$ can be estimated\footnote{The number of orbits required to inspiral by a distance $\Delta r$ can be estimated as $\Delta N \sim \Delta r /(\dot{r}_\mathrm{GW} T)$, where $\dot{r}_\mathrm{GW}$ is the rate of inspiral due to GW emission. We also assume $m_2 \ll m_1$.} as $\Delta N \approx (r/R_\mathrm{ISCO})^{3/2}$, where $R_\mathrm{ISCO}$ is the ISCO of the central BH $m_1$. This number $\Delta N$ is parametrically much smaller than $N_\mathrm{crit} \approx r/r_\mathrm{acc}$ for accretion by a realistic BH, except at very large radii. Thus, we expect that taking into account the inspiraling of the accreting BH, the asymmetry in the DM spike induced by the accretion process will not be sufficient to substantially affect the formalism presented here.

\section{Quasicircular inspirals with accretion} \label{sec:circ_acc}

Gravitational waveforms from small mass ratio inspirals are sensitive to the environments in which they evolve and therefore serve as a potential probe of dark matter.
Many previous works have focused on the effects of dynamical friction (e.g.~\cite{Antonini_2011,Dosopoulou_2017,Edwards:2019tzf,Kavanagh:2020cfn}), while some have explored accretion in the context of scalar clouds~\cite{Baumann:2021fkf,Boudon_2022} or unbound DM~\cite{shapiro2023spikes}. In the context of CDM spikes, \citet{Yue:2017iwc} argued that accretion in a static laminar flow model is of negligible importance. However, in light of the recent work by \citet{Kavanagh:2020cfn}, it was shown that dynamical friction is greatly suppressed when accounting for a dynamic CDM spike distribution. Thus, the previous estimates would place accretion as a comparable effect to that of dynamical friction. 

In this section, we incorporate the prescription for accretion onto the companion into our evolution method for the binary. We discuss the evolution equations and our numerical method for solving them; and then present our simulated results.

\subsection{Evolution equations and numerical methods}
In \cref{sec:intro}, we presented the formalism for the evolution of a binary system in the absence of accretion. Operating within the post-Newtonian approximation, we accounted for gravitational wave radiation at the 2.5 order PN, but omitted the 1PN and 2PN conservative terms and worked with Newtonian orbital energy and angular momentum for the regimes of interest. The binary was described through its semi-major axis $a$ and orbital eccentricity $e$, while the spike through its distribution function $f(\calE)$. The orbital element $(a, e)$ would then evolve according to \cref{eq:da_dtdedt}, with changes in the orbital energy $E$ and angular momentum $L$ driven by gravitational wave emission and dynamical friction. The evolution of the distribution function would be described by \cref{eq:df_dt}.

Because we now also include the effects of accretion, we will alter this prescription to account for the change in mass of the companion $m_2$ through \cref{eq:accretion_rates2}; the force of accretion in \cref{eq:accretion_force}; the depletion of the distribution function from \cref{eq:accretion_feedback}; and finally alter the evolution equations for the binary to account for the change in mass of the companion, given by \cref{eq:da_dt,eq:de_dt}. 

In the quasi-circular regime, this prescription reduces to the following system of differential equations,
\begin{equation} \label{eq:coevolution}
\begin{aligned}
    \dot{m}_2 &= \rho_\mathrm{DM} \sigma_\mathrm{BH} u\, \mathcal{C}_\mathrm{m}\,,\\
    \dot{r}_2 &= r_2 \bigg[ \frac{\dot{E}_\mathrm{tot}}{Gm_1m_2/(2r_2)} -\frac{\dot{m}_2}{m} \bigg]\\
    \dot{E}_\mathrm{tot} &= -\frac{32 G^4}{5c^5} \frac{\mu^2 m^3}{r_2^5} -4\pi G^2 m_2^2 \frac{{\rho_\mathrm{DM}}}{u} \mathcal{C}_\mathrm{DF} \\
    &\qquad\qquad- \rho_\mathrm{DM} \sigma_\mathrm{BH} u^3\, \mathcal{C}_\mathrm{acc}\\
  \dot{f}  &= \frac{\Delta_{+}f(\calE) - \left(p_\mathrm{DF}+p_\mathrm{acc} \right) f(\calE) }{T} \,,
\end{aligned}
\end{equation}
where $r_2$ is the binary separation. Here $\rho_\mathrm{DM}$, $\sigma_\mathrm{BH}$, and the various prefactors $\mathcal{C}$ (some of which are functions of the companion's orbital velocity) are evaluated at each separation $r_2$. In the total energy loss $\dot{E}_\mathrm{tot}$, the first term is due to gravitational wave radiation, the second is the dynamical friction, and the third is the accretion. The first term in the distribution function evolution $\dot{f}$ comes from particle re-distribution due to dynamical friction, the second is the dynamical friction depletion, the third is the accretion depletion. The pre-factor $p_\mathrm{DF}$ is the dynamical friction pre-factor of \cref{eq:DF}, and $p_\mathrm{acc}$ is the accretion probability of \cref{eq:accretion_probability}. The DM density $\rho_\mathrm{DM}$ is obtained by integrating the distribution function $f(\calE)$ over velocities.

With this scheme, we can self-consistently evolve any dressed binary system within the assumptions stated in \cref{sec:intro}. To solve these equations, we have produced Python code that utilizes Ralston's method, a 2nd order Runge-Kutta. In it, the binary is described by the two evolving scalars $r_2$, $m_2$, and a distribution function $f(\calE, t)$ that is evaluated on a dense grid of energies $\calE$, based on the \texttt{HaloFeedback} implementation~\cite{HaloFeedback}. The integration evolves with dynamically chosen timesteps $\Delta t$ that are tuned with an Euler method to avoid large relative changes in the scalars or nonphysical values to the distribution function.

\subsection{Numerical simulation results}
First, we will qualitatively describe the behavior of the binary based on \cref{eq:coevolution}. Let us make this exploration through a comparison of three systems; a binary in vacuum space (``Vacuum''); a ``dressed'' binary in the dark matter spike where only the effects of dynamical friction are included similarly to \citet{Kavanagh:2020cfn} (hereafter ``DF''); and another dressed binary where both dynamical friction and accretion are included (``DF + AC'').\footnote{The comparison between the ``DF'' and ``DF+AC'' systems is not only representation of the improvement in the calculation of dressed binaries, but also a comparison of the impact of the companion's nature on the evolution of the system. In the case where the companion is a compact object without an event horizon, standard CDM models do not predict any accretion.} In practice, the ``DF+AC'' system is the same as \cref{eq:coevolution}, while for ``DF'' we take $\mathcal{C}_m = \mathcal{C}_\mathrm{acc} = p_\mathrm{acc}(\calE) = 0$, and for ``Vacuum'' also take $\mathcal{C}_\mathrm{DF} = \dot{f}(\calE) = 0$.

Based on \cref{eq:coevolution}, we expect that the vacuum case (with only the GW radiation switched on) to evolve in such a way that the binary's separation decreases monotonically. When dynamical friction is included, the decrease is accelerated due to its dissipatory nature, represented by a negative contribution to $\dot{E}_\mathrm{tot}$. Finally when accretion is switched on, although more nuanced, the inspiral is faster still. By isolating the contribution of accretion
\begin{equation}
\frac{\dot{r}_2}{r_2} \propto -\left(\mathcal{C}_\mathrm{acc} +\frac{q}{2} \mathcal{C}_\mathrm{m} \right)\,.
\end{equation}
we see how the first term, the energy dissipation by accretion of momentum, and the second, the deepening of the potential due to mass increase, both accelerate the inspiral. However, the latter is negligible here as it is being suppressed by the small mass ratio $q$ while the two prefactors ($\mathcal{C}_\mathrm{acc}$, $\mathcal{C}_m$) are of similar order.

\begin{table}[tb]
  \centering
  \begin{tabular}{c c c c c}
      \toprule
      \vspace*{0.3em}
      $m_1$ [$M_\odot$] & $m_2$ [$M_\odot$] & $\rho_6$ [$M_\odot/\mathrm{pc}^3$] & $\rho_\rmsp$ [$M_\odot/\mathrm{pc}^3$] & $\gamma_\rmsp$\\
      $10^4$ & 10 & $10^{16}$ & 1.1 & 7/3 \\
      \botrule
  \end{tabular}
  \caption{\textbf{The dark dress benchmark system of this study.} The columns indicate the black hole masses chosen for a mass ratio that allows for measurable feedback \cite{Kavanagh:2020cfn}, and initial dark matter spike parameters within expected ranges~\cite{Cole:2022yzw}.
  \label{tab:benchmarks}}
\end{table}

\begin{figure}[tb]
  \hspace*{-0.85cm}
  \includegraphics[width=\columnwidth]{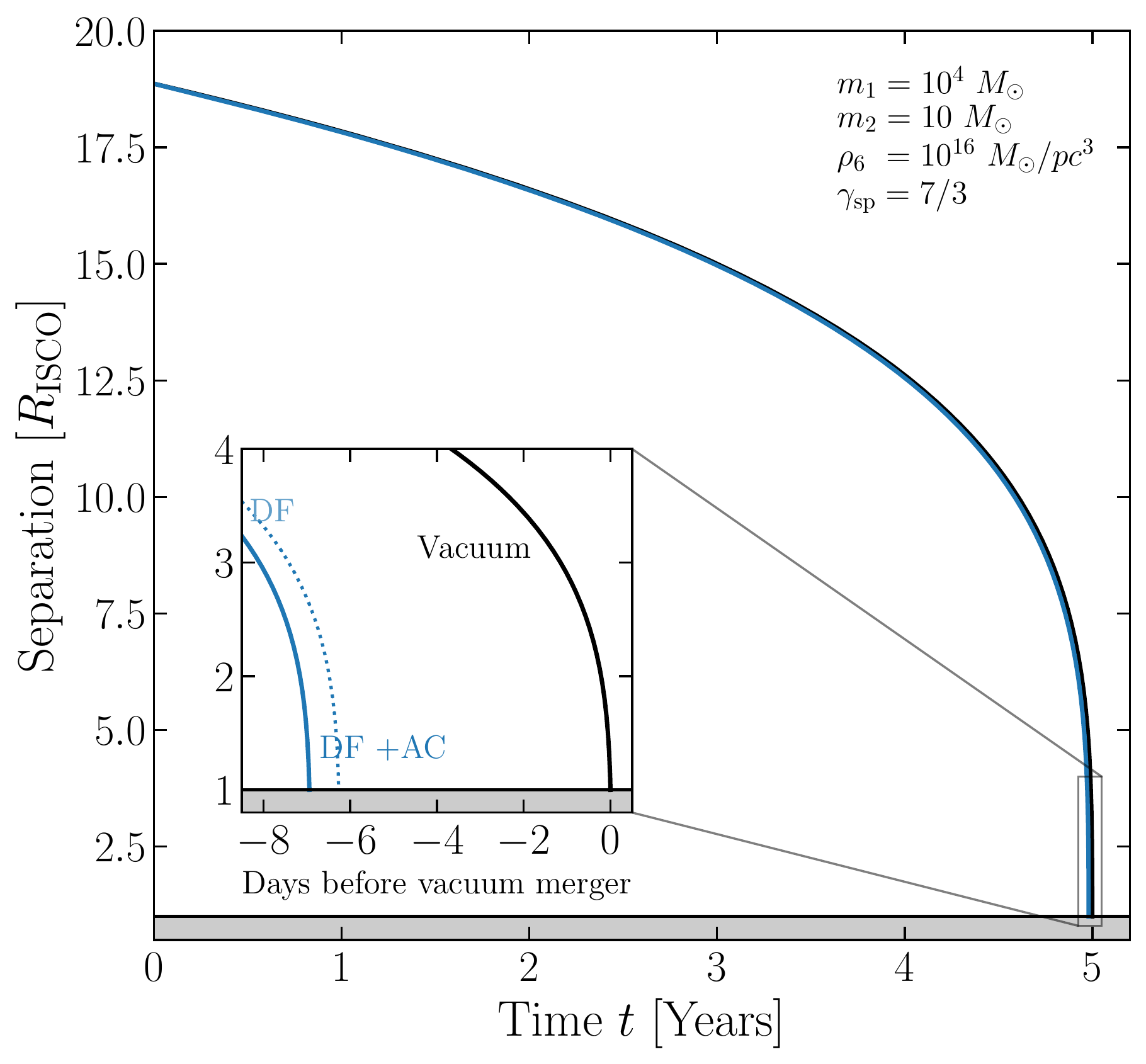}
  \caption{\textbf{Evolution of the binary separation $r_2$.} A benchmark parameter system is examined in presence and absence of a spike. The vacuum case (solid black) is the slowest to inspiral, while the ``DF+AC'' case (solid blue) is the fastest. The ``DF'' case (dashed blue) is in between the two.  Accretion will noticeably accelerate the inspiral and cause the binary to merge sooner.
  \label{fig:separation_evolution}
  }
\end{figure}

In \cref{fig:separation_evolution}, we plot the evolution of the binary separation $r_2$ for the three systems described above. The benchmark parameters from \cref{tab:benchmarks} represent a typical dressed IMRI system of two black holes. Inspired by the LISA mission~\cite{LISA,Seoane:2021kkk}, we choose to evolve the system in such a way that it would merge within 5 years if it were in vacuum space (roughly $r_{2,i\,} = 18.87 R_\mathrm{ISCO}$).\footnote{The choice of initial conditions is not straightforward in the study of environmental effects. Energy and mass exchange during past inspirals may alter the environment. The above setup is equivalent to the assumption that the black hole companion has instantly appeared in a circular orbit inside a dark matter spike to form an IMRI with the central black hole.} We indeed recover the expected behavior and find that the three systems which start at the same initial separation, merge approximately after 5 years, but at different times. Specifically, the ``DF+AC'' and ``DF'' systems merge about 7 and less than 6 days sooner than the vacuum case respectively.

Additionally, we find that the increase in the companion's mass is minimal (about $\delta x=$ 0.25\% or 0.025 $M_\odot$), and occurs predominantly at larger separations. This is due to the longer inspiral timescales at larger separations, which allow for more accretion.\\

\begin{figure}[t!]
  \hspace*{-0.85cm}
  \includegraphics[width=\columnwidth]{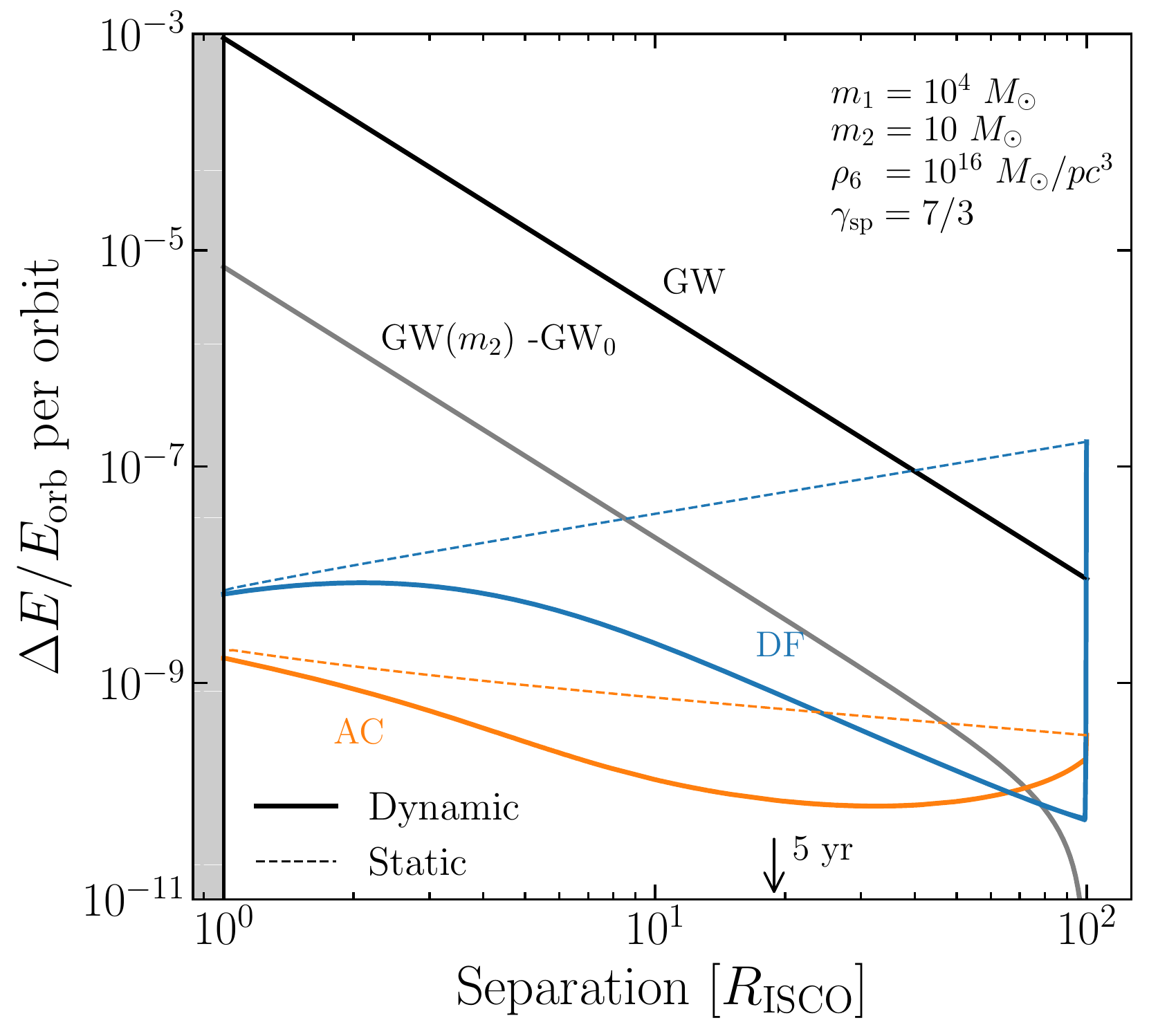}
  \caption{\textbf{The fraction of the orbital energy that is lost per orbit at different separations.} The dashed lines correspond to a spike in absence of feedback. The force of accretion overwhelms that of dynamical friction near the initial separations and the change in gravitational waves due to the increase in the companion's mass takes over both of them. The sharp initial features of the blue (and orange) lines are short, transient periods of spike depletion. These arise from not choosing to simulate the companion entering an unperturbed spike but instead appearing already inside of it at a later configuration. \cite{Coogan:2021uqv,Cole:2022ucw}.
  \label{fig:energy_evolution}
  }
\end{figure}

\paragraph{\bf Energy loss curves.}
To better understand the behaviour of the binary, we look at the evolution of the energy loss terms with the separation. In \cref{fig:energy_evolution} we show the fraction of the orbital energy that is lost per orbit for each mechanism in the case where all forces are active and feedback is active (solid lines). We also show the case where the spike is present, but feedback is not (dashed lines). We choose the same benchmark system parameters, but we initialize the binary at $r_2 = 100 R_\mathrm{ISCO}$.

As expected, gravitational wave radiation remains the driving force of the inspiral at most separations. Furthermore, dynamical friction is larger than accretion, but only by a small amount. This is in contrast to previous claims that dynamical friction is the dominant spike-induced force in the system~\cite{Yue:2017iwc}. This difference is due to dynamical friction feedback which dampens the strength of that force~\cite{Kavanagh:2020cfn} towards, what we empirically observe to be, a comparable scale to that of accretion. For small separations, when feedback is negligible, the losses of the dynamically evolved systems tend to their unperturbed counterparts, while at the largest a transient period of DM redistribution arises due to the simulation's unphysical initialization. Otherwise, we find that dynamical friction follows a broken power law which is described by the ``shell model'' of \citet{Coogan:2021uqv}. This is interesting as the result should formally deviate from this prediction when accounting for accretion in this regime; either through the increased inspiral rate $\dot{r}_2$ or the accretion's feedback rate $p_\mathrm{acc}f(\calE)/T$. However, the relative effect is negligible. Finally, we note that the increase in the companion's mass also has a substantial effect on the gravitational wave radiation which scales with $m_2^2$. The gray line depicts this by showing the change in power to this emission as the mass increases.\\

\paragraph*{\bf Sensitivity to initial conditions.}

In this system, the initial conditions of the binary and the spike are coupled through the feedback mechanism. \citet{Cole:2022ucw} have shown that under the effects of dynamical friction, the DM density will converge to a universal profile, regardless of the starting separation of the binary, as long as the feedback timescales are longer than the inspiral timescale. 
This removes some of the sensitivity to initial conditions in most regimes. 

In our setup, we have identified a similar behaviour. First, when the system is initialized the accretion energy loss will quickly alter and deplete alongside dynamical friction. Then, there is a transient stage of slower depletion before turning into an enhancement. The separation where this switch occurs seems to be related to the initialization separation and is further away the further the initial separation. Finally, at the turning point when accretion begins to increase again, we observe convergence to a universal curve similar to that in \cite{Cole:2022ucw}.\\

\paragraph*{\bf GW signal dephasing.}

To quantify the size of the effect induced by dark matter, we calculate the difference between the number of gravitational wave cycles accumulated during an inspiral in vacuum and in the presence of a spike. A system with more sources of energy dissipation, other than GW radiation, will experience an accelerated orbital decay and subsequently accumulate a smaller number of cycles before it merges. In the case where the frequency monotonically increases in time, we define the number of GW cycles between two frequencies $f_1$ and $f_2$ as,
\begin{equation}
  N(f_1, f_2) = \int_{f_1}^{f_2} f \, \bigg(\frac{\mathrm{d}f}{\mathrm{d}t} \bigg)^{-1} \mathrm{d}f \,,
\end{equation}
where the gravitational wave frequency is twice the orbital frequency  in the quadrupole approximation~\cite{Maggiore:2007ulw}, and $\mathrm{d}f/\mathrm{d}t$ is the total derivative of the frequency with respect to time, accounting for the change in separation and mass. We focus on the cycles remaining until the binary merges, which we approximate through the frequency associated with the Innermost Stable Circular Orbit (ISCO) and define $N_{\to c}(f) \equiv N(f, f_c = 2 f_\mathrm{ISCO})$. 

We define the GW dephasing as the difference in remaining cycles (in radians) between the vacuum $N_V(f)$ and spike $N_{S}(f)$ case respectively:
\begin{equation}
    \label{eq:dephasing}
    \Delta \Phi_{\to c} (f) = 2\pi \int_f^{f_c} f' \, \bigg( \frac{\mathrm{d}f}{\mathrm{d}t}\bigg|^{-1}_{V} -\frac{\mathrm{d}f}{\mathrm{d}t}\bigg|^{-1}_{S} \bigg) \mathrm{d}f' \,.
\end{equation}
Alternatively, using \cref{eq:da_dt}, we can express this in terms of the semi-major axis $a(f)$ up to $\mathcal{O}(q)$:
\begin{equation}
    \label{eq:dephasing_a}
    \Delta \Phi_{\to c}(f) = -\pi G m_1 \int_{a}^{a_c} \left( \frac{f_\mathrm{V} m_{2,V}}{\dot{E}_V} -\frac{f m_{2}}{\dot{E}} \right) \frac{\mathrm{d}a}{a^2} \,,
\end{equation}
where $a_c = R_\mathrm{ISCO}$. Here, $\dot{E}_V$ is the energy loss rate of gravitational waves in the vacuum system; $m_{2,V}$ the constant companion's mass; and $\dot{E}$ the total energy loss in the spike including GWs and two forces. Even in absence of the forces, the presence of a mass accretion rate alters $m_2(a)$ and thus the GW emission, leading to a third source of dephasing.

\begin{figure}[tb]
    \hspace*{-0.85cm}
    \includegraphics[width=\columnwidth]{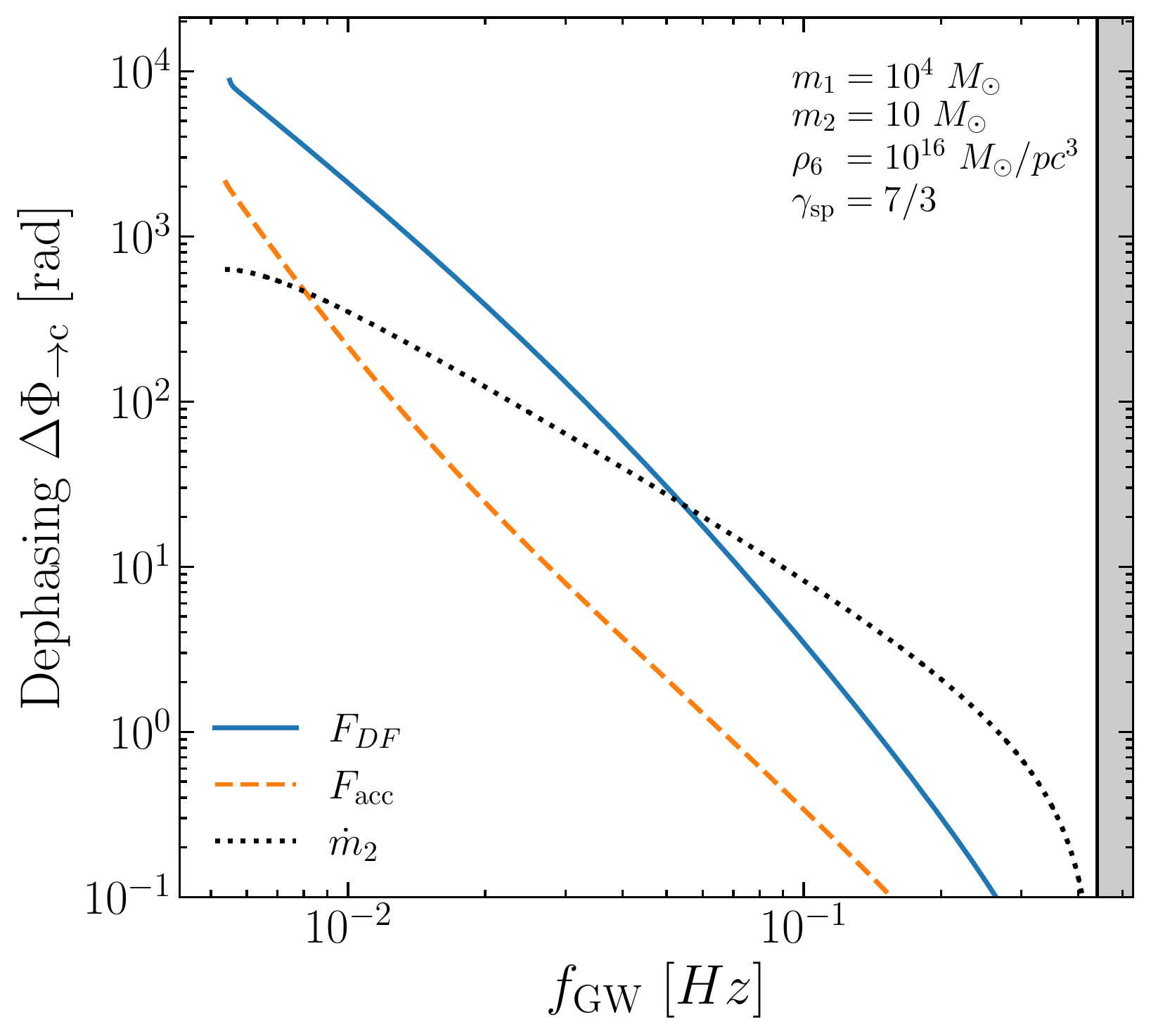}
    \hspace*{-0.85cm}
    \includegraphics[width=\columnwidth]{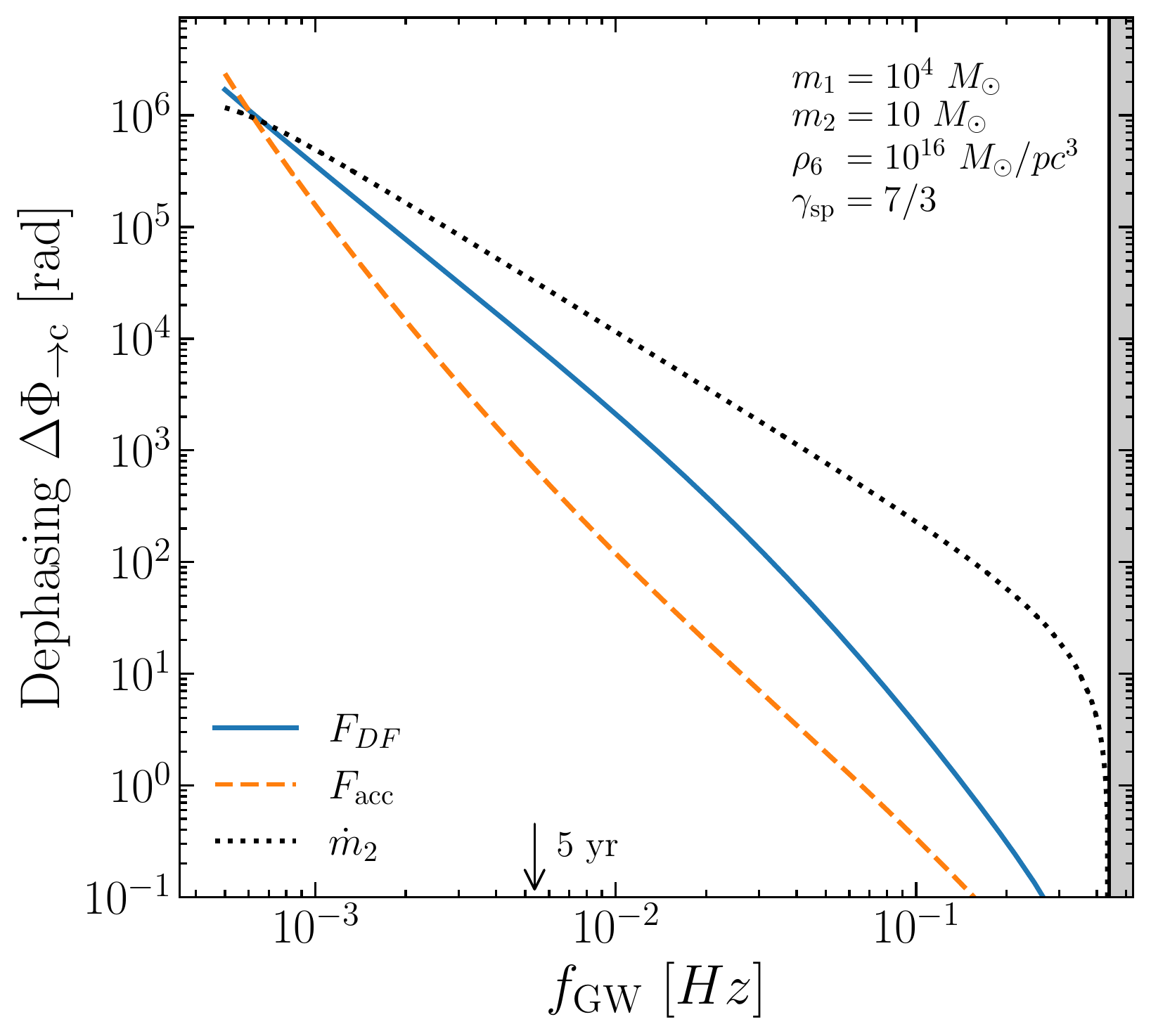}
    \caption{\textbf{Dephasing of the gravitational wave signal from different sources.} We show the difference in the gravitational wave cycle count (starting at some frequency $f_\mathrm{GW}$ until the merger at $f_c = 2f_\mathrm{ISCO} \approx 0.44$~Hz) between a vacuum inspiral and a system with a dynamic dark matter spike. We assume the benchmark astrophysical system with parameters given in \cref{tab:benchmarks}. The upper panel shows results starting 5 years from the merger, while the lower panel shows a longer inspiral, starting at a larger separation, which shows the comparable effects of dynamical friction and accretion forces.
    \label{fig:dephasing}
    }
\end{figure}

In \cref{fig:dephasing}, we show this dephasing for the benchmark system described above, where $f_c \approx 0.44$ Hz. Specifically, the lines indicates the dephasing of the gravitational wave as a function of the wave's frequency $f_\mathrm{GW}$ between the vacuum case and the spike case. Each line shows the contribution to to this dephasing by a particular effect. The solid blue line is that induced by dynamical friction alone, while the dashed orange and dotted black lines are due to the two different effects of particle accretion, $F_\mathrm{acc}$ and $\dot{m}_2$ respectively.

The dephasing induced by accretion is a large effect in terms of cycles lost, and becomes dominant in lower and higher frequencies through $F_\mathrm{acc}$ and $\dot{m}_2$ respectively. The dephasing effects attributed to the two forces induces by the spike are comparable to each other and dynamical friction takes over at higher frequencies with its dephasing $\Delta \Phi_{\to c} (f)$ approximating that of the static case in absence of feedback \cite{Coogan:2021uqv}. This system was initialized at $r_2 = 100 R_\mathrm{ISCO}$ and we show the dephasing from this point in the lower panel of \cref{fig:dephasing}. In the upper panel, we show the dephasing starting 5 years before merger (at $f_\mathrm{5\,yr} \approx 0.005$ Hz) which shows a dephasing of $10,000$ radians from dynamical friction, $700$ from the force of accretion and $30,000$ from the change in the gravitational wave emission.

The final contribution is much larger than the rest due to gravitational wave's relative potency in extracting energy from the system. As the mass of the companion increases slightly from $\dot{m}_2$, the already GW-dominated system inspirals slightly faster primarily by the increase of GW radiation, and less so by the suppressed deepening of the energy potential. Thus it draws fewer cycles compared to a vacuum counterpart where no mass was ever accreted. We reason that this effect to be of the order $\delta x \Phi_V$ where $\Phi_V$ is the total phase of the vacuum case and $\delta x$ (typically $<1\%$) is the fractional change of the companion's mass. Because this is a large number for EMRIs and IMRIs, this effect tends to be large. A dephasing comparison with a best-fit vacuum system is left for future discussion.

\section{Inspirals with non-zero eccentricity} \label{sec:eccentricity}

The preparation for gravitational wave searches of dressed EMRIs has usually been carried in the quasi-circular regime~\cite{Cardoso_2021} due to the circularization strength of gravitational radiation. However, environments add additional level of nuance to the problem. For dark matter spikes, \citet{Yue_2019} suggested that dynamical friction eccentrifies the orbit at large separation when accounting for a toy model of the spike, ignoring feedback and the phase space distribution of the particles. Conversely, when these effects are included, \citet{Becker_2022} and \citet{Li2} found that the spike induces a moderate net circularization instead.

In this section, we present a formalism for the treatment of inspirals with any orbital eccentricity beyond the static case, meaning that we include feedback from dynamical friction and that of accretion discussed in \cref{sec:accretion}. The layout is as follows, we will first describe the evolution equations and numerical method, expanding the feedback of dynamical friction from previous work~\cite{Kavanagh:2020cfn} to the case of eccentric inspirals. Then, we will briefly explore the co-dependence of the evolution of the binary and spike in a toy system, before finally examining the complete case. 

\subsection{Evolution equations and eccentric feedback} \label{sec:feedback}

When the binary is allowed to have a non-zero orbital eccentricity at any point during its inspiral the evolution equations of \cref{eq:coevolution} are no longer valid and must be supplemented. In particular, the binary separation $r_2$ is replaced with the semi-major axis $a$ of the orbit, and the orbital eccentricity $e$ is introduced as a new evolving parameter. We do this through equations \cref{eq:da_dt,eq:de_dt}, including angular momentum loss along with energy loss. The new equations will take the form:
\begin{equation} \label{eq:coevolution_ecc}
\begin{aligned}
  \dot{a} &= -a \bigg[ \frac{\dot{E}_\mathrm{tot}}{E} +\frac{2\dot{\bar{m}}_2 - \dot{m}_2}{m} \bigg]\,,\\
  \dot{e} &= -\frac{1-e^2}{e} \bigg[ \frac{\dot{E}_\mathrm{tot}}{2E} + \frac{\dot{L}_\mathrm{tot}}{L} +\frac{\dot{\bar{m}}_2 - \dot{m}_2}{m} \bigg]\,, \\
  \dot{f}  &= \frac{\Delta_{+}f(\calE) - \left(p_\mathrm{DF}+p_\mathrm{acc} \right) f(\calE) }{T} \,,
\end{aligned}
\end{equation}
where the total loss rates $\dot{E}_\mathrm{tot}$ and $\dot{L}_\mathrm{tot}$ are obtained as the sum of GW radiation (\cref{eq:GWlosses}) or environmental effects (substituting \cref{eq:DF,eq:accretion_force} in \cref{eq:forceLosses}), and $\dot{\bar{m}}_2$ is a orbit-weighted mass accretion rate.  Explicitly, these are
\begin{equation} \label{eq:coevolution_ecc2}
  \begin{aligned}
    \dot{E}_\mathrm{tot} &= -\frac{32 G^4}{5c^5} \frac{\mu^2 m^3}{a^5}\frac{1 +\frac{73}{24}e^2 +\frac{37}{96}e^4 }{(1-e^2)^{7/2}} \\
    &\quad\, -4\pi G^2 m_2^2 \frac{{\rho_\mathrm{DM}}}{u} \mathcal{C}_\mathrm{DF} - \rho_\mathrm{DM} \sigma_\mathrm{BH} u^3\, \mathcal{C}_\mathrm{acc} \,, \\
    \dot{L}_\mathrm{tot} &= -\frac{32 G^{7/2}}{5c^5} \frac{\mu^2 m^{5/2}}{a^{7/2}}\frac{1 +\frac{7}{8}e^2 }{(1-e^2)^{2}} \\
    &\quad\, -4\pi G^2 m_2^2 \frac{{\rho_\mathrm{DM}}}{u} \mathcal{C}_\mathrm{DF} - \rho_\mathrm{DM} \sigma_\mathrm{BH} u^3\, \mathcal{C}_\mathrm{acc} \,, \\
    \dot{m}_2 &= \rho_\mathrm{DM} \sigma_\mathrm{BH} u\, \mathcal{C}_\mathrm{m}\,, \qquad  \dot{\bar{m}}_2 = \frac{r_2}{a} \dot{m}_2 \,. \\
  \end{aligned}
\end{equation}
The environmental quantities are then integrated over an entire elliptical orbit through \cref{eq:averageEnergy} before injecting them in \cref{eq:coevolution_ecc}.

Finally, the distribution function evolution has the same form as in \cref{eq:coevolution}, but each of the probabilities must now be evaluated following a general elliptical orbit. Below, we elaborate on each feedback term.\\

\paragraph{\bf Accretion feedback.}
In the case of accretion, we have already derived the term $p_\mathrm{acc}(\calE)$ for general orbits which can be used as is from \cref{eq:accretion_probability}. Interestingly, the form of the final integration over $\theta$ allows us to gain some intuition on the effect of increasing eccentricity to the probability of particle accretion. Specifically, for elliptical orbits $p_\mathrm{acc}(\calE)$ takes the form of an effective average over $\theta$ of the circular case. This allows for more particles to be accreted at higher eccentricities.\\

\paragraph{\bf Dynamical friction feedback}

Unlike the novel accretion feedback derived in \cref{eq:accretion_probability}, when dynamical friction feedback was first conceived by \citet{Kavanagh:2020cfn} it was done so in the circular orbit limit. Thus, in this part we will summarize the formalism and adapt the calculation to the case of elliptical orbits.

Following a similar logic to the one presented in \cref{sec:massdepletion}, we aim at building the distribution function differential of \cref{eq:df_dt} and re-calculating it for the elliptical case. Because the work transferred by dynamical friction is not lost, particles that scatter off the companion will receive a boost in energy $\Delta\calE$ and thus be displaced to other orbits. If the probability of receiving such a boost $P_\calE(\Delta\calE)$ is known, then the change in the amount of particles at a given energy $\calE$ sees a contribution from those that were boosted away from that energy (depletion) and those that are boosted into it from other states (replenishing). This can be written as:
\begin{align} \label{eq:Ndf}
  \Delta N(\calE) &= - N(\calE) \int\displaylimits P_\calE(\Delta \calE) \,\mathrm{d}\Delta\calE \\&\quad +\int\displaylimits N(\calE -\Delta\calE) P_{\calE -\Delta\calE}(\Delta \calE) \,\mathrm{d}\Delta\calE\,. \nonumber
\end{align}
Or equivalently, in terms of the distribution function differential
\begin{align}
  &\dot{f}(\calE) = - f(\calE) \frac{1}{T} \int\displaylimits P_\calE(\Delta \calE) \,\mathrm{d}\Delta\calE \\ &+\frac{1}{T} \int\displaylimits_{\Delta\calE_\mathrm{min}}^{\Delta\calE_\mathrm{max}} \bigg( \frac{\calE}{\calE -\Delta\calE} \bigg)^{5/2} f(\calE -\Delta\calE) P_{\calE -\Delta\calE}(\Delta \calE) \,\mathrm{d}\Delta\calE \nonumber \,,
\end{align}
where the limits come from $\Delta \calE(b) = -2u^2 (1+b^2/b_{90}^2)^{-1}$~\cite{Kavanagh:2020cfn} for $b = b_\mathrm{min}$ and $b_\mathrm{max}$ at the periapsis or apoapsis of the orbit.

The probability $P_\calE(\Delta\calE)$ is that of a particle with energy $\calE$ receiving a boost $\Delta\calE$. We calculate this from the reduced density of states $g'$ of particles with energy $\calE$ that receive this boost, meaning $P_\calE(\Delta\calE) = g'/g$. We note that size of the boost is a function of the impact parameter $b$ of the particle's orbit $\Delta\calE(b)$. With this, then, the reduced density of states is given by
\begin{equation}
   g'=\iint{ \delta(\calE(r, v) -\calE) \, \delta(\Delta\calE(b) -\Delta\calE)\, \mathrm{d}^3 \\ \bm{r}\,\mathrm{d}^3\, \textbf{v}}\,.
\end{equation}
The details of this calculation follow closely that of \cref{sec:massdepletion} and are presented in \cref{app:df_probability}, where we rectify a small error in the original derivation of \citet{Kavanagh:2020cfn} in choosing the torus volume element (see \cref{app:deformed_torus}). Ultimately, the probability takes the form,
\begin{align}
\begin{split}
  P_\calE(\Delta\calE) &= \frac{4\pi G^2 m_2^2 \sqrt{2}}{g(\calE)\,\Delta\calE^2} \\
  &\mkern-36mu \times \iint  \frac{r_2}{u^2} \sqrt{\Psi(r_2)(1 -\frac{b_*}{r_2}\cos\phi) -\calE} \,\mathrm{d}\phi \,\mathrm{d}\theta \,.
\end{split}
\end{align}
Just as with \cref{eq:accretion_probability}, the role of the $\theta$ integration is understood as an effective averaging of the circular case over the orbit, and hence we utilize the \texttt{HaloFeedback} implementation~\cite{HaloFeedback} to evaluate each sub-integral over $\phi$ with special elliptic functions.\\

\paragraph{\bf Understanding feedback and eccentricity.}

To understand the interplay between feedback and elliptical orbits, let us evaluate a particular numerical experiment. Specifically, we consider the time-evolution of a black hole binary of constant mass and locked in place. In practice, this is achieved by setting $\dot{a}=\dot{e}=0$ in \cref{eq:coevolution_ecc}, only including the $\dot{f}$ terms for dynamical friction and accretion. Although this is not a realistic description of the system which will inevitably alter its orbital parameters, it provides some understanding of how the spike distribution is altered over short timescales.

\begin{figure}[tb]
  \includegraphics[width=\columnwidth]{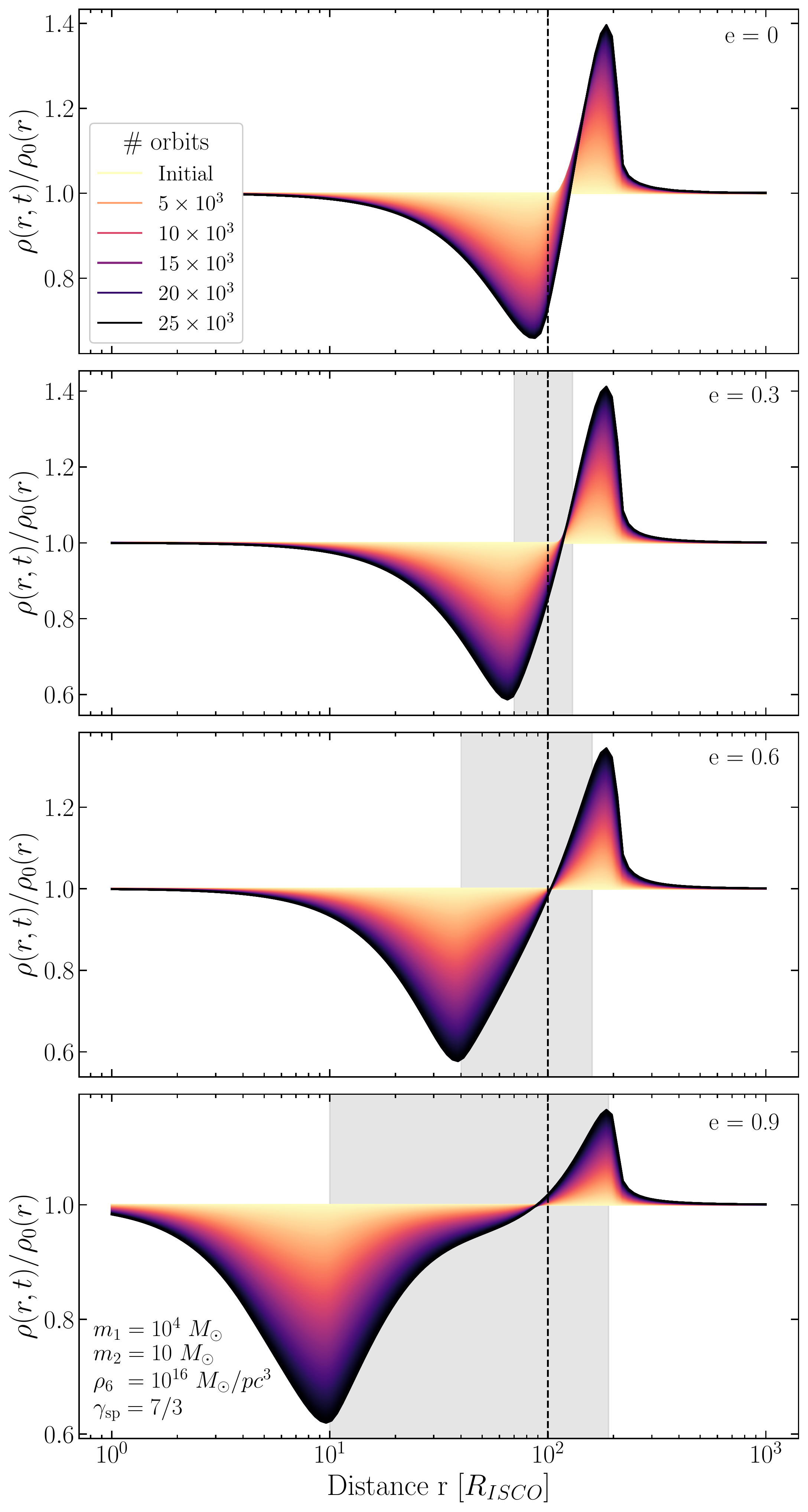}
  \caption{\textbf{Changes to the density profile due to dynamical friction and accretion.}
  Each binary is on a \textit{fixed} orbit with constant orbital elements and $a = 10^2 R_{ISCO}$ simulated for $25,000$ orbits. The dashed vertical line represents the semi-major axis $a$ of the systems, and the gray area shows the radial extent of the orbit, i.e.~$r_2 \in [a(1-e), a(1+e)]$. 
  }
  \label{fig:evolvingProfiles}
\end{figure}

The four panels in \cref{fig:evolvingProfiles} show heatmaps of the time-evolution of the density profile of four binaries with benchmark parameters and on orbits of different eccentricity. Specifically, their orbital parameters are $a = 10^3 \, R_\mathrm{ISCO}$ and $e \in \{0, 0.3, 0.6, 0.9\}$ from top to bottom. Each system is evolved for up to $25,000$ orbits or approximately $223$ years and their profile is extracted at various reference times and reported with a different color. Specifically, we plot the relative density at each time, meaning the current density over the initial $\rho(t)/\rho(t=0)$ for various values of distance from the center $r$.

In the top panel, where the dark matter halo is allowed to evolve in response to the orbit of a circular IMRI, we recover the behavior reported by \citet{Kavanagh:2020cfn}. That is, the halo exhibits two regions, one depleted at lower separations and another denser; as predicted by the functional form of \cref{eq:Ndf} which dominates feedback. Particles of lower energies will gain energy and on average be displaced towards wider orbits. In fact, these two regions are common in all other panel configurations as well. As we increase the orbital eccentricity the companion is allowed to access more shells in the spike (as illustrated by the gray regions) and increasingly so towards the center, scaling with $r_\mathrm{min} = a(1 -e)$. Indeed, we see that the largest depletion occurs at $r_\mathrm{min}$. These depleted particles are displaced over the wide range of radii covered by the eccentric orbit, ultimately making the excess density at large radii less pronounced than in the circular case.

\subsection{Numerical simulation results}
We will now present the results of our numerical simulation. We investigate a more realistic binary system which is allowed to evolve fully, and we report the evolution of the density profile, orbital eccentricity as well as of the second mode dephasing from quadrupole radiation.\\

\paragraph{\bf Density profile.}

\begin{figure}[tb]
  \centering
  \hspace{-0.8cm}
  \includegraphics[width=\columnwidth]{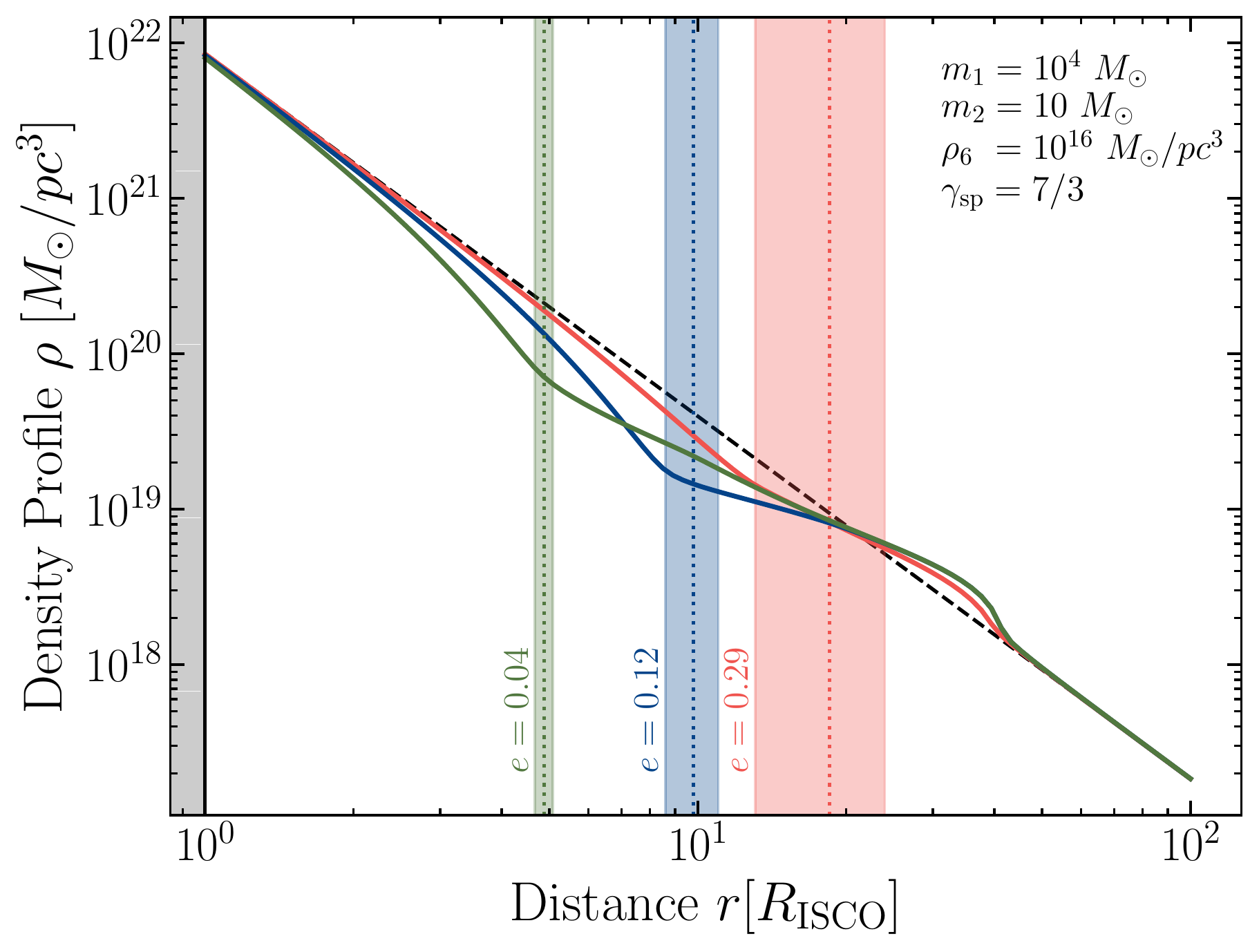}
  \caption{\textbf{Density profile during binary evolution.} An eccentric IMRI with $e_0=0.3$ is evolved until it merges. Snapshots of the density profile are shown after some time has elapsed where the binary has reached eccentricities $e\approx[0.29, 0.12, 0.04]$ respectively. The colored regions correspond to the range of radii covered by the companion's eccentric orbit, while the vertical dotted line is its semi-major axis. The black dashed line is the initial unperturbed DM density profile.
  }
  \label{fig:density_snapshots_eccentric_complete}
\end{figure}

In \cref{fig:density_snapshots_eccentric_complete}, we simulate a benchmark system with initial eccentricity $e_0 = 0.3$ and at a large semi-major axis. We show three snapshots of the density profile at various times during its evolution and report the system's orbital eccentricity at that point. The orbital eccentricity decreases as the inspiral progresses not only due to gravitational wave radiation but also due to dynamical friction and accretion. However, the environment's effect is largely suppressed in the presence of feedback. In line with the previous numerical experiment, the first snapshot exhibits a small depletion at distances smaller than those available to the companion's orbit and an increased density at larger distances. However, at later times when the binary progresses through the previously depleted areas it dynamically replenishes matter behind its path leading to profile snapshots that resemble broken power-laws aligning with the orbit's periapsis. This behaviour recovers the quasi-circular case in \cite{Kavanagh:2020cfn} where the density profile breaks at a distance equal to the binary's separation $r_\mathrm{min}(e=0) \to r_2$. Finally, at the end of the evolution the profile will rest at a similar but steeper power-law.\\

\begin{figure}[tb]
  \hspace*{-1cm}
  \includegraphics[width=0.95\columnwidth]{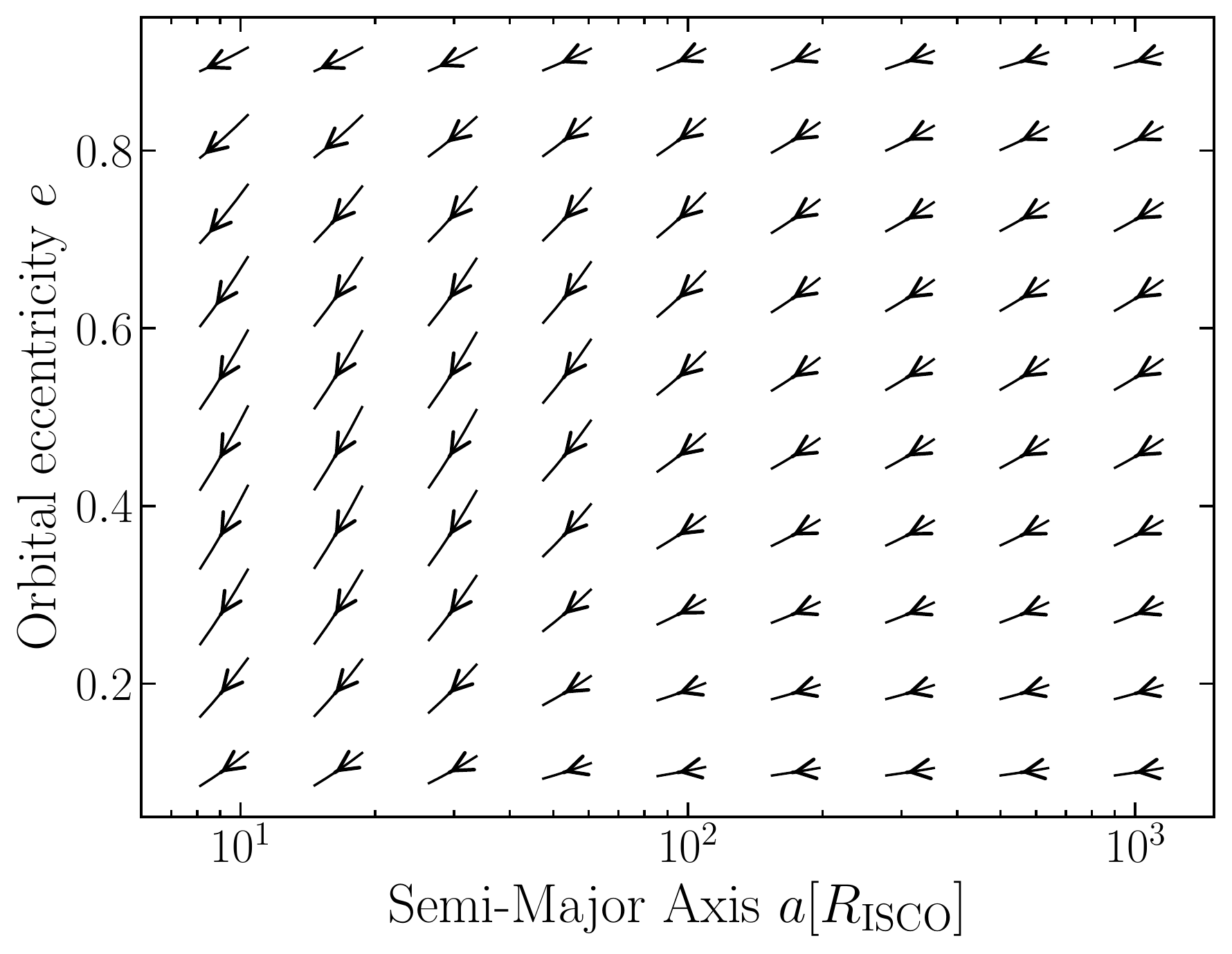}
  \caption{\textbf{Direction of orbital eccentricity evolution in dynamically evolved spikes.} A benchmark system is allowed to evolve for a short period for a range of initial orbital parameters $a_0, e_0$, each corresponding to one point. The vector at each point is the direction of the relative derivative $\dot{e}/\dot{a}$. Circularization is expected across the whole range of orbital configurations shown here.
  \label{fig:eccentricity_points}
  }
\end{figure}

\paragraph{\bf Orbital circularization.} In the presence of the spike, the system's eccentricity evolves under the influence of dynamical friction and mass accretion as well as gravitational wave emission. In \cref{fig:eccentricity_points} we depict a vector map corresponding to $\dot{e}/\dot{a}$, the direction of change in eccentricity corresponding to the orbit's semi-major axis shrinking. Each arrow is for a benchmark parameter system's first few orbits after a binary is formed in a static profile. We find that for all points, the eccentricity tends to decrease and the orbit to shrink. This is expected at smaller separations where gravitational wave radiation dominates the energy loss \cite{Maggiore:2007ulw} and in the regime where inspiral timescales are shorter than those of feedback~\cite{Becker_2022}. In the latter regime, although the spike contributes to circularization of the orbit, the large force of dynamical friction increases $\dot{a}$ and accelerates the inspiral, leading to marginally larger eccentricities than vacuum binaries but still smaller compared to the static treatement \cite{Becker_2022}. We find that this picture does not change when including accretion. In addition, in regions where feedback is substantial, we find that the spike's contribution to the eccentricity remains in the direction of circularization.\footnote{We find orbit circularization is always the case except momentarily for a negligible increase in eccentricity during the initial spike depletion at the beginning of certain simulations. This is an artifact of the initial spike re-distribution at very low eccentricities and circularization quickly ensues again.} In the regime at small separations, where GW emission dominates the effects from DF and accretion are greatly suppressed and thus only marginally contribute to the evolution. Thus the system's eccentricity evolution is driven primarily by gravitational wave emission.\\

\paragraph{\bf Orbital dephasing.} Finally, we look at the interplay of orbital eccentricity and gravitational dephasing. While this is straightforward in the circular regime where gravitational wave radiation is primarily emitted at the second mode $f_\mathrm{GW} = 2 f_\mathrm{orb}$, elliptic systems will emit gravitational waves of varying power at different frequencies~\cite{Maggiore:2007ulw}. To make a first comparison, we will restrict ourselves to the second mode of the emission, and thus also effectively compare the difference in the orbital cycles drawn by the binary itself. 

\begin{figure}[tb]
  \hspace*{0.25cm}
  \includegraphics[width=\columnwidth]{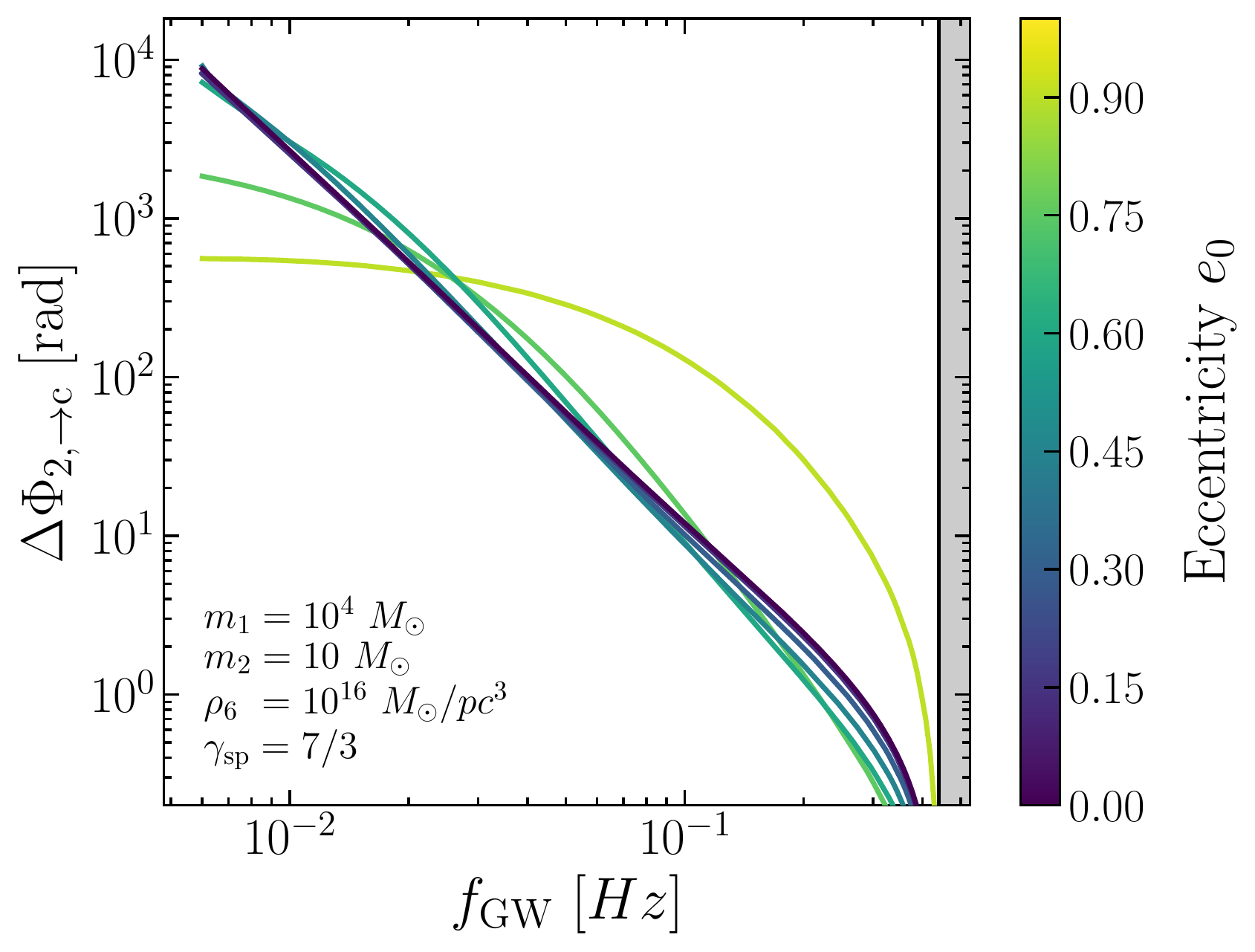}
  \caption{\textbf{Dephasing for different initial eccentricities.} Most of the curves follow closely the shape of the quasi-circular case. For large eccentricities and high frequencies, accretion takes over as the dominant contribution to dephasing, as the growing mass of the secondary enhances the GW emission. We assume the benchmark astrophysical system with parameters given in \cref{tab:benchmarks}.
  \label{fig:dephasing_ecc}
  }
\end{figure}

In \cref{fig:dephasing_ecc}, we show the dephasing for our benchmark system (\cref{tab:benchmarks}) initialized at the same semi-major axis but with different orbital eccentricities.\footnote{Note that this initial configuration gives rise to quadrupole radiation roughly inside the LISA sensitivity.} Remarkably, we find that as the eccentricity increases up to $e_0 = 0.6$, the dephasing is well within 40\% of the circular case for all the frequency range. Smaller orbital eccentricities are even more similar; for the case of $e_0=0.3$ the dephasing shows less than a 10\% difference compared with the circular case. In fact, we find that most of the change in the dephasing is accumulated at very low frequencies when spike depletion is dominant, and at higher frequencies where the accumulated mass due to accretion enhances GW emission. Finally, we find that for higher values of initial eccentricity, the curve becomes increasingly different, acquiring most of its dephasing at larger frequencies.

\section{Conclusions and Discussion} \label{sec:conclusion}

In this work we have presented two advancements in the study of intermediate and extreme mass ratio inspirals inside dark matter (DM) spikes: DM accretion from the spike onto the companion, and the interplay between orbital eccentricity and spike.

We have derived the contribution of particle accretion to the black hole (BH) companion, taking into account an evolving particle distribution, and provided a semi-analytic formalism that self-consistently captures this effect on the evolution of BH binaries as well as the surrounding DM distribution. In the formalism, this takes the form of an additional force $F_\mathrm{acc}$ acting on the companion,\footnote{This novel force of accretion serves as the equivalent to Chandrasekhar's dynamical friction for orbits which are absorbed by the BH, as opposed to orbits which are deflected by the BH at larger impact parameters.}  a time-dependent mass of the companion $\dot{m}_2$, and a change in the distribution function $\dot{f}_{acc}$ as particles are depleted by accretion. To verify our prescription, we present a series of numerical $N$-body experiments, described in a companion paper~\cite{BetterSpikesII}, which accurately recover all of our analytical predictions. 

Utilizing this refined description, we have shown that particle accretion leads to an accelerated inspiral due to the increased rate of energy loss, and thus to a large dephasing effect on the gravitational waveforms that generally cannot be ignored. We have found that the dephasing effect due to the accretion force is dominant to that of dynamical friction at low frequency (large separations). At high frequency (small separations), we have also found that a second effect becomes relevant: the growing mass of the secondary BH enhances the rate of inspiral due to GW emission, leading to a dephasing with respect to the vacuum case.

While this paper was in preparation, \citet{PhysRevD.108.124062} presented an analysis of particle accretion onto the companion in DM spikes. Their prescription also takes into consideration the increase in the companion's mass and includes a term for mass conservation in the spike's distribution function. The main difference between our prescriptions is in the regime of their validity. For implementing the effects of accretion, \citet{PhysRevD.108.124062} considers the increase in mass of the companion through dimensional analysis and its back-reaction on the orbit. In this toy model, the rate of mass change is only proportional to the density profile at the position of the companion, and otherwise independent of the distribution function. In our formalism, this is equivalent to setting the mass prefactor in \cref{eq:accretion_ratesproper} to 1. This is only valid when the companion moves much faster than the DM particles $u \gg v$.
Indeed, we find that our accretion feedback formalism recovers their reported results in the limit where $u \gg v$ and when ignoring the distribution of all possible particle encounters.
Finally, because mass is not conserved, the equations of motion must be altered, which we find to give rise to a novel back-reaction force with its own prefactor (\cref{eq:accretion_ratesproper2}) that corresponds to the accumulating accretion of momentum from the absorbed particles.

In the second part of this paper, we have analyzed the interplay between orbital eccentricity and dynamic DM distributions for the first time. We did this by extending the feedback model of \citet{Kavanagh:2020cfn} to the case of general orbits and included it in our simulation prescription. By including this effect, we extend the domain of applicability of the eccentric formalism presented by \citet{Becker_2022} (valid only in the static case) to the case of intermediate mass ratio inspirals with mass ratio $q = m_2/m_1 \leq 10^{-5}$, where feedback is generally strong at some point in the inspiral. 

We find that when accounting for dynamical friction and accretion from the spike, each effect contributes to the circularization of the orbit. Specifically, the circularization rate with respect to the semi-major axis of the orbit is found to be larger than in the static case and equivalent to that of GW emission alone. This is because of the large suppression of the environmental effects when feedback is strong. Finally, we investigate the dephasing of the emitted waveforms with respect to the vacuum case for quadrupole emission. We find that for systems entering the LISA band with initial eccentricities up to $e_0 \approx 0.6$, the dephasing closely follows the shape of the circular case. Conversely, for larger eccentricities the dephasing deviates considerably and is suppressed at low frequencies but considerably accumulates towards higher frequencies.

In this work, we have focused on feedback on the DM spike due to close two-body encounters with the secondary (which give rise to dynamical friction) and due to accretion. In the companion paper~\cite{BetterSpikesII}, we present the results of $N$-body simulations, which suggest an additional mechanism of feedback on the spike. This ``stirring" mechanism arises from the motion of the DM particles in the time-dependent potential of the binary, and can impact the DM distribution at radii larger than the orbital separation. Future work should aim to incorporate this ``stirring mechanism" into the framework described here. 
Future work should also explore how accretion-related effects would alter the measurability of the DM spike properties in future GW detectors. Indeed, the extension of feedback effects to include both accretion and eccentricity which we present here is a crucial step towards future searches for dark matter dephasing with gravitational waves.

\section*{Acknowledgements}
TK acknowledges the support 
of the University of Amsterdam through the APAS-2020-GRAPPA MSc scholarship, and of Sapienza Università di Roma for support at the ``EuCAPT Workshop: Gravitational Wave probes for black hole environments".
BJK acknowledges funding from the Ram\'on y Cajal Grant RYC2021-034757-I and from the \textit{Consolidaci\'on Investigadora} Project \textsc{DarkSpikesGW}, reference CNS2023-144071, both financed by MCIN/AEI/10.13039/501100011033 and by the European Union ``NextGenerationEU"/PRTR.
GB gratefully acknowledges the support of the Italian Academy of Advanced Studies in America and of the Department of Physics of Columbia University, where this work was finalised.

We acknowledge Santander Supercomputing support group at the University of Cantabria who provided access to the supercomputer Altamira at the Institute of Physics of Cantabria (IFCA-CSIC), member of the Spanish Supercomputing Network, for performing simulations.

We also thank Niklas Becker for useful comments on the paper. 
\appendix

\section{Volume element for a deformed torus} \label{app:deformed_torus}
When calculating the scattering and absorption probabilities for dynamical friction and accretion respectively, we restrict the density of states integration to a ring around the companion. As the companion completes an orbit around the central black hole a torus is formed on its path. To integrate inside this geometry we parameterize the system with the torus coordinates
\begin{align}
    x &= (r_2(\theta) + b\cos\phi)\cos\theta\,, \\
    y &= (r_2(\theta) + b\cos\phi)\sin\theta\,, \\
    z &= b\sin\phi\,,
\end{align}
where the major radius is the binary's separation $r_2(\theta)$, the minor radius is the impact parameter of the encounter $b$, $\theta \in [0, 2\pi]$ is the azimuthal angle of the axis around which the volume is drawn -- the binary's true anomaly -- and $\phi \in [0, 2\pi]$ is the angle tracing the inner cross section of the torus. The translation of the differential $
\mathrm{d}^3\bm{r}$ to the torus coordinates is given by the determinant of the transformation's Jacobian. The Jacobian is written as
\begin{equation}
    \scalebox{0.92}{$\displaystyle
\textbf{J} =
\begin{bmatrix}
  \cos\phi\cos\theta & 
    -b\sin\phi\cos\theta & 
    r'_2(\theta)\cos\theta-R\sin\theta \\[1ex]
  \cos\phi\sin\theta & 
    -b\sin\phi\sin\theta & 
    r'_2(\theta)\sin\theta +R\cos\theta \\[1ex]
  \sin\phi & 
    b\cos\phi & 
   0
\end{bmatrix}\,,
$}
\end{equation}
where the term $r'(\theta)$ is a partial derivative over $\theta$, and $R=r_2~+~b\cos\phi$. The determinant is evaluated as $\left|\left|\textbf{J}\right|\right| = b \, (r_2 +b\cos\phi)$, which has the same analytical form to that of a proper torus of constant major radius. 

\section{Accretion probability} \label{app:accretion_probability}

In \cref{sec:massdepletion}, we present a formalism for self-consistently removing particles from the spikes in response to accretion by an  inspiraling black hole. We do this by calculating the fraction of the spike's density of states (DoS) corresponding to particles that satisfy the condition for accretion onto the companion, $b \leq b_\mathrm{acc}$. Whilst the total available DoS is trivially evaluated as an integration over all possible spatial and velocity particle configurations (see \cref{eq:DoS0}), the calculation of the restricted DoS $g'$ is more nuanced.

To define this special DoS, we enforce the accretion condition as an additional restriction to \cref{eq:DoS0}, writing:
\begin{align}
  g' &= \iint \delta\big(\calE -\Psi(r)+v^2/2\big) H\big(b-b_\mathrm{acc}\big) \,\mathrm{d}^3\bm{r} \,\mathrm{d}^3\mathbf{v}\,.
\end{align}
We solve this integration in two parts, evaluating most of the velocity component utilizing the spherical symmetry, before working on the spatial component in a special coordinate system that follows the companion. Here, we invoke the secular evolution of the inspiral to justify a calculation over a single, constant orbit of the companion.

In spherical coordinates, the isotropic velocity distribution allows us to express the differential as \linebreak $\mathrm{d}^3\mathbf{v}=v^2 \,\mathrm{d}v \,\mathrm{d}\cos\gamma \,\mathrm{d}\alpha$, taking $\gamma$ as the angle between $\mathbf{v}$ and $\bm{u}$ without loss of generality. Then, aiming to evaluate over the magnitude of the velocity $v$ we express the Dirac delta as $\delta\big(\calE -\Psi(r)+v^2/2\big) = \delta(v-v_*)/v_*$, where $v_*(\calE, r) = \sqrt{2\big(\Psi(r) -\calE \big)}$; thus finding
\begin{align}
  g' &= \iint H\big(b-b'_\mathrm{acc}\big) v_* \,\mathrm{d}\cos\gamma \,\mathrm{d}\alpha\,\mathrm{d}^3\bm{r}\,,
\end{align}
where $b'_\mathrm{acc}$ is now a function of $v_*$ through its relation with \cref{eq:b_acc}. Our accretion condition restricts the spatial volume to that of particles with an impact parameter smaller than a critical value as they follow the companion. For this, we switch to the coordinates of a deformed torus that follows companion with major and minor radius $r_2(\theta)$ and $b_\mathrm{acc}$ respectively; thus taking $\mathrm{d}^3\textbf{r}~=~b(r_2~+~b\cos\phi)\, \mathrm{d}b\, \mathrm{d}\theta\, \mathrm{d}\phi$ derived in \cref{app:deformed_torus}.\footnote{Here, we rectify a technical error from \cite{Kavanagh:2020cfn} in the Jacobian of the transformation; fortunately the results converge at first order.} Because of the open integration around the true anomaly $\theta$ the result is, for the first time, robust for all eccentricities in the inspiral. We find,
\begin{equation}
  g' = \iint v_* b(r_2 +b\cos\phi) \,\mathrm{d}\cos\gamma \,\mathrm{d}\alpha\, \mathrm{d}b\, \mathrm{d}\theta\, \mathrm{d}\phi\,,
\end{equation}
where $b \in [0, b'_\mathrm{acc}]$, $\cos\gamma \in [-1, 1]$, $\alpha \in [0, 2\pi]$, $\theta \in [0, 2\pi]$, and $\phi \in [0, 2\pi]$. Then, due to the relatively small size of the accretion radius, we approximate at first order in $b/r_2$, taking the distance to the particle $r$, as $r \to r_2$ within the integrand. This allows us to redefine $v_* = \sqrt{2\big(\Psi(r_2) -\calE \big)}$, and thus we can integrate over the impact parameter $b$ and angles $\alpha$, $\phi$ trivially,
\begin{equation}
  g' = 2\pi \iint v_* r_2 \sigma' \,\mathrm{d}\cos\gamma \, \mathrm{d}\theta\,,
\end{equation}
where $\sigma' = \pi b'^2_\mathrm{acc}$ is the accretion cross-section of the companion for particles with velocity $v_*$. Substituting \cref{eq:c_section} for the cross section above, and $\cos\gamma = x$ we find
\begin{equation}
  g' = 2\pi^2 r_\mathrm{acc}^2 \zeta \iint v_* r_2 \bigg( 1 +\frac{r_s}{r_\mathrm{acc}}\frac{c^2}{u^2 +v_*^2 -2uv_*x} \bigg) \,\mathrm{d}x \, \mathrm{d}\theta\,,
\end{equation}
and after integrating over $x$,
\begin{equation}
  g' = 4\pi^2 r_\mathrm{acc}^2 \zeta \int v_* r_2 \bigg( 1 +\frac{r_s}{r_\mathrm{acc}}\frac{c^2}{2uv_*} \ln\frac{u+v_*}{|u-v_*|} \bigg) \, \mathrm{d}\theta\,. \label{eq:g_prime}
\end{equation}
Finally, combining \cref{eq:g_prime,eq:bayesProb} we arrive at the probability of a particle with specific energy $\calE$ to be accreted by the companion as
\begin{equation}
    p_\mathrm{acc}(\calE) = \frac{4\pi^2}{g(\calE)} \zeta r_\mathrm{acc}^2 \int r_2 \bigg(v_* +\frac{r_s}{r_\mathrm{acc}} \frac{c^2}{2u} \ln \frac{u+v_*}{|u-v_*|} \bigg) \, \mathrm{d}\theta\,.
\end{equation}

We can see that the probability is in part proportional to the square of the accretion radius, as expected, mimicking the geometric cross-section of the companion. This final integration over the true anomaly $\theta$ of the companion captures the change in this reduced density of states in response to the different orbital separations for non-zero eccentricities. For this reason, and because it is not trivial, we perform this integration numerically. Additionally, we note that a logarithmic pole appears in the integrand, but it does not appear to alter our results.

If we were to take the limit of $v \ll u$, it reduces to the expression,
\begin{equation}
  p_\mathrm{acc}(\calE) = \frac{4\pi^2}{g(\calE)} \zeta r_\mathrm{acc}^2 \int v_* r_2 \bigg(1 +\frac{r_s}{r_\mathrm{acc}} \frac{c^2}{u^2} \bigg)\, \mathrm{d}\theta\,,
\end{equation}
proportional to the cross-section of the companion. We identify this as the probability that recovers the mass extracted by the companion in the laminar flow case (see our discussion above \cref{eq:spike_mass_laminar}), nonetheless we find that it does not meaningfully alter our results beyond the first order correction.\footnote{As this work was in preparation, \citet{PhysRevD.108.124062} recently used this version of the probability in their analysis of accretion by the companion in the laminar flow.}

\section{Scattering probability}
\label{app:df_probability}
We aim to calculate the probability $P_\calE(\Delta\calE)$ of a particle with specific energy $\calE$ to be scattered by the companion to a new energy $\calE +\Delta\calE$. This is the derivation of \citet{Kavanagh:2020cfn} adapted to our notation and extended beyond circular orbits. We start with the reduced density of states for this case,
\begin{equation}
  g'=\iint{ \delta(\calE(r, v) -\calE) \, \delta(\Delta\calE(b) -\Delta\calE)\, \mathrm{d}^3 \bm{r}\,\mathrm{d}^3 \textbf{v}}\,.
\end{equation} \vspace{1em}
\noindent Then we substitute the delta functions,
\begin{equation}
  g' = \frac{G^2 m_2^2}{\Delta\calE^2} \iint \frac{\delta(v -v_*)}{v_*} \frac{\delta(b -b_*)}{b_*u^2}\,\mathrm{d}^3 \,\bm{r} \,\mathrm{d}^3 \textbf{v}  \,,
\end{equation}
where $v_*=\sqrt{2(\Psi(r) -\calE)}$ and $b_*=b_{90}\sqrt{\frac{2}{1+q}\frac{u^2}{\Delta\calE} -1}$ are the roots of the respective surfaces inside the Dirac deltas. After evaluating the integration over all particle velocities $v$ we obtain
\begin{equation} \label{eq:df_probs_condition}
  g' = \frac{4\pi G^2 m_2^2}{\Delta\calE^2} \!\!\!\!\! \int\displaylimits_{|\textbf{r}| = r_\mathrm{cut}}^{r_{\calE}} \!\!\!\!\! v_* \frac{\delta(b -b_*)}{b_*u^2} \,\mathrm{d}^3 \textbf{r} \,,
\end{equation}
where the two integration limits are: the maximum allowed radius of a particle at a given state $r_\calE=Gm_2/(\calE)$, and the minimum distance $r_\mathrm{cut} = Gm_2/(\calE +u_\mathrm{cut}^2 /2)$ which ensures scattering with only particles slower than $u_\mathrm{cut} \to u$. The remaining delta function restricts the spatial volume to a cross-section of radius $b$ which trails the separation $r_2$ of the companion. This forms a (deformed) torus of major radius $r_2(\theta)$ and minor radius $b_*(\Delta\calE)$, whose volume element is $d^3\textbf{r} = b(r_2 +b\, \cos\phi)\, \mathrm{d}b\, \mathrm{d}\theta\, \mathrm{d}\phi$ (see \cref{app:deformed_torus}). Substituting these and evaluating the Dirac delta we find,
\begin{equation}
  g' = \frac{4\pi G^2 m_2^2}{\Delta\calE^2} \iint v'_* \frac{r_2 +b_* \cos\phi}{u^2} \,\mathrm{d}\phi\, \mathrm{d}\theta \,,
\end{equation}
which is non-zero for $b_\mathrm{min} \leq b_* \leq b_\mathrm{max}$. The new velocity quantity $v_*'=\sqrt{2(\Psi-\calE)}$ where $\Psi = \Psi(r)$ the gravitational potential of the binary, evaluated at $r = \sqrt{r_2^2 +b^2_* +2r_2b_*\, \cos\phi}$. At leading order in $b_*/r_2$ we drop the $b_* \cos\phi$ term and take $r \approx r_2/(1 -b_*\cos\phi /r_2)$, hence
\begin{align}
\begin{split}
  g' &= \frac{4\pi G^2 m_2^2 \sqrt{2}}{\Delta\calE^2} \times \\
  &\iint \frac{r_2}{u^2} \sqrt{\Psi(r_2)\left(1 -\frac{b_*}{r_2}\cos\phi\right) -\calE} \,\mathrm{d}\phi\, \mathrm{d}\theta \,.
\end{split}
\end{align}
Finally, the integration over $\phi$ is performed over half a circle (hence a factor 2) at the region $\alpha_1 = \cos^{-1}\big[ \min\{(r_2 -r^2_2/r_\calE)/b_*, 1\} \big]$, $\alpha_2 = \cos^{-1}\big[ \max\{(r_2 -r^2_2/r_\mathrm{cut})/b_*, -1\} \big]$ in order to enforce the condition on $r$ from \cref{eq:df_probs_condition}. The result is evaluated using special elliptic functions as in~\cite{Kavanagh:2020cfn}. The integration over $\theta$ remains to be evaluated numerically around the companion's orbit $r_2(\theta)$.

\bibliography{references}

\end{document}